\documentclass[aps,prb,reprint,superscriptaddress]{revtex4-2}
\usepackage{graphicx}
\usepackage{dcolumn}
\usepackage{bm}
\usepackage{hhline}
\usepackage{tabularx}
\usepackage{tablefootnote}
\usepackage{gensymb}
\usepackage{hyperref}
\usepackage{xfrac}
\hypersetup{
    colorlinks=true,
    linkcolor=blue,
    filecolor=blue
    urlcolor=blue,
    citecolor=blue,
}

\begin{document}

\title{Evidence for incommensurate antiferromagnetism in nonsymmorphic UPd$_{0.65}$Bi$_2$}

\author{S.~Mishra} \email[]{sm335@rice.edu}
\altaffiliation{Present address: Department of Physics and Astronomy, Rice University, Houston, Texas, 77005, USA}
\affiliation{Los Alamos National Laboratory, Los Alamos, New Mexico, 87545, USA}
\author{C. S. Kengle}\affiliation{Los Alamos National Laboratory, Los Alamos, New Mexico, 87545, USA}

\author{J. D. Thompson}\affiliation{Los Alamos National Laboratory, Los Alamos, New Mexico, 87545, USA}

\author{A. O. Scheie}\affiliation{Los Alamos National Laboratory, Los Alamos, New Mexico, 87545, USA}

\author{S. M. Thomas}\affiliation{Los Alamos National Laboratory, Los Alamos, New Mexico, 87545, USA}

\author{F. Ronning}\affiliation{Los Alamos National Laboratory, Los Alamos, New Mexico, 87545, USA}

\author{P. F. S. Rosa}\affiliation{Los Alamos National Laboratory, Los Alamos, New Mexico, 87545, USA}

\date{\today}

\begin{abstract}
The intersection between nonsymmorphic symmetry and electronic correlations has emerged as a platform for topological Kondo semimetallic states and unconventional spin textures. Here we report the synthesis of nonsymmorphic UPd$_{0.65}$Bi$_2$ single crystals and their structural, electronic, magnetic, and thermodynamic properties. UPd$_{0.65}$Bi$_2$ orders antiferromagnetically (AFM) below $T_N\simeq$ 161~K as evidenced by a sharp cusp in magnetic susceptibility, a second-order phase transition in specific heat, and an upturn in electrical resistivity, which suggests an incommensurate AFM structure that deviates from the A-type magnetism typically observed in this class of materials. Across $T_N$, Hall effect measurements reveal a change from electron-dominated to hole-dominated transport, which points to a sharp reconstruction in the electronic structure at $T_N$. Upon further cooling, a first-order transition is observed at $T_1 \simeq 30~$K in magnetic susceptibility and heat capacity but not in electrical resistivity or Hall measurements, which indicates a small change in the AFM structure that does not affect the electronic structure. Our specific heat data reveal a small Sommerfeld coefficient ($\gamma \simeq$13 mJmol$^{-1}$K$^{-2}$), consistent with localized 5$f$ electrons. Our results indicate that UPd$_{0.65}$Bi$_2$ hosts weak electronic correlations and is likely away from a Kondo semimetallic state. Low-temperature magnetization measurements show that the AFM structure is remarkably stable to 160~kOe and does not undergo any field-induced transitions. Neutron diffraction and magnetization experiments at higher fields would be valuable to probe the presence of unconventional spin textures.
\end{abstract}

\maketitle

\subsection{Introduction}
Quantum materials that harness topologically protected functionalities are sought after due to their potential applications in quantum information sciences and spintronics \cite{Qi, Biao, Gilbert, He, Smejkal}.
In non-interacting systems, nonsymmorphic \cite{Schoop, Schoop2018, Watanabe, Po} and non-centrosymmetric \cite{Bian1, Bian} space group symmetries have played a crucial role in the realization of topological gapless states \cite{bradlyn, cano}. In strongly correlated electron materials, the intersection between electronic correlations and crystalline symmetries has recently emerged as a promising avenue to uncover novel topological phenomena \cite{Dzsaber2, Asaba, Onuki, Kang}.
 Intermetallic $4f$ compounds are a particularly tunable platform due to the interplay between Kondo interactions, Ruderman–Kittel–Kasuya–Yosida (RKKY) interactions, and the local crystalline electric field environment. For example, in the Kondo limit, recent theory developments reveal that nonsymmorphic tetragonal space group 129 ($P4/nmm$) is a favorable structure for Weyl-Kondo semimetallic states with nodes pinned to the Fermi energy, which may be realized in CeRh$_2$Ga$_2$ \cite{Chen2022}. In the 4$f$-localized limit, the same space group (SG) also hosts CeAgBi$_{2}$, which displays topological spin textures due to its locally noncentrosymmetric structure \cite{Simeth2024}.

Synthesizing novel $f$-electron-based materials that crystallize in nonsymmorphic space groups may therefore offer a route towards strongly correlated non-trivial topological states of matter wherein the underlying crystalline symmetry of the lattice cooperates with strong electronic correlations. Here we investigate a relatively underexplored platform: $5f$-based nonsymmorphic materials. In particular, the family of compounds U$MX_2$  ($M=$ transition metal, $X =$ pnictogen) exhibits some of the essential ingredients for correlated topological behavior: nonsymmorphic symmetry (SG 129), Kondo semimetallic behavior, and tunable magnetic ground states \cite{Kaczorowski}. For instance, in antimonide U$M$Sb$_2$, a ferromagnetic Kondo semimetallic state was observed for $M =$ Co, Cu, Ag, and Au, whereas antiferromagnetic order was observed for $M = $ Ni, Ru, and Pd \cite{Kaczorowski, KACZOROWSKI1992333, IKEDA200462}. In the case of bismuthides ($X =$ Bi), UAuBi$_2$ is a ferromagnetic Kondo semimetal \cite{PhysRevB.92.104425}, whereas UCuBi$_2$ and UNiBi$_2$ are antiferromagnetic \cite{KACZOROWSKI1992333}. Recently, UAgBi$_2$ has been synthesized and shown to display four different magnetic phases as a function of temperature, which points to competing interactions and multiple nearly degenerate states \cite{Gabriel2024}.

In this paper, we present the structural, electronic, magnetic, and thermodynamic properties of UPd$_{0.65}$Bi$_2$. Our results show that UPd$_{0.65}$Bi$_2$ undergoes an antiferromagnetic transition at $T_N =$ 161~K followed by a first-order transition at 30 K, which is likely related to a change in the magnetic structure. Magnetic susceptibility and resistivity measurements suggest an incommensurate magnetic structure. Hall effect measurements reveal a drastic change in the Fermi surface at $T_N$ from electron-dominated to hole-dominated transport. Our specific heat measurements reveal a small Sommerfeld coefficient, $\gamma \simeq$13 mJmol$^{-1}$K$^{-2}$, which points to local 5$f$ moments. Our results indicate that UPd$_{0.65}$Bi$_2$ hosts weak electronic correlations and is likely away from a Kondo semimetallic state. In contrast to other members in this family, the AFM structure of UPd$_{0.65}$Bi$_2$ is remarkably stable to 160~kOe and does not undergo any field-induced transitions. Neutron diffraction  and magnetization experiments at higher fields are required to probe the presence of unconventional spin textures.

\subsection{Experimental details}
Single crystals of UPd$_{0.65}$Bi$_2$ were obtained from a bismuth self-flux technique \cite{Rosa2019}. Uranium and palladium pieces in a 1:1 ratio were first arc melted in a water-cooled Cu hearth prior to the growth.  UPd and Bi pieces were then loaded  in a 1:10 ratio into an alumina crucible and sealed under vacuum in a quartz ampoule. The reagents were heated to 950$^\circ$C and held at this temperature for 24 h before being slowly cooled at 4$^\circ$C/h to 550$^\circ$C. The excess Bi flux was then removed by centrifugation, and plate-like crystals were obtained in the crucible.

The crystallographic structure of UPd$_{0.65}$Bi$_2$ was verified at room temperature by a Bruker D8 Venture single-crystal x-ray diffractometer equipped with Mo K-$\alpha$ radiation. Raw data were processed with the Bruker SAINT software, including multi-scan absorption correction. Initial crystallographic models were obtained via the intrinsic phasing method in SHELXT. A least-squares refinement was performed with SHELXL2018 \cite{Sheldrick2015acta}. Elemental analysis of our single crystals was performed using energy-dispersive x-ray spectroscopy in a commercial scanning electron microscope.

Magnetic susceptibility and magnetization measurements on a UPd$_{0.65}$Bi$_2$ single crystal were carried out in a Quantum Design (QD) SQUID magnetometer. Magnetization measurement to 160~kOe are performed in a QD PPMS using the vibrating sample magnetometer option. Specific heat measurements were performed using a QD calorimeter that utilizes a quasi-adiabatic thermal relaxation technique. Electrical resistivity and Hall measurements were performed using a standard four-point configuration method and an ac resistance bridge.

\subsection{Results}
\subsubsection{Crystal Structure Determination}

\begin{figure}
	\centering
   \includegraphics[width=\columnwidth]{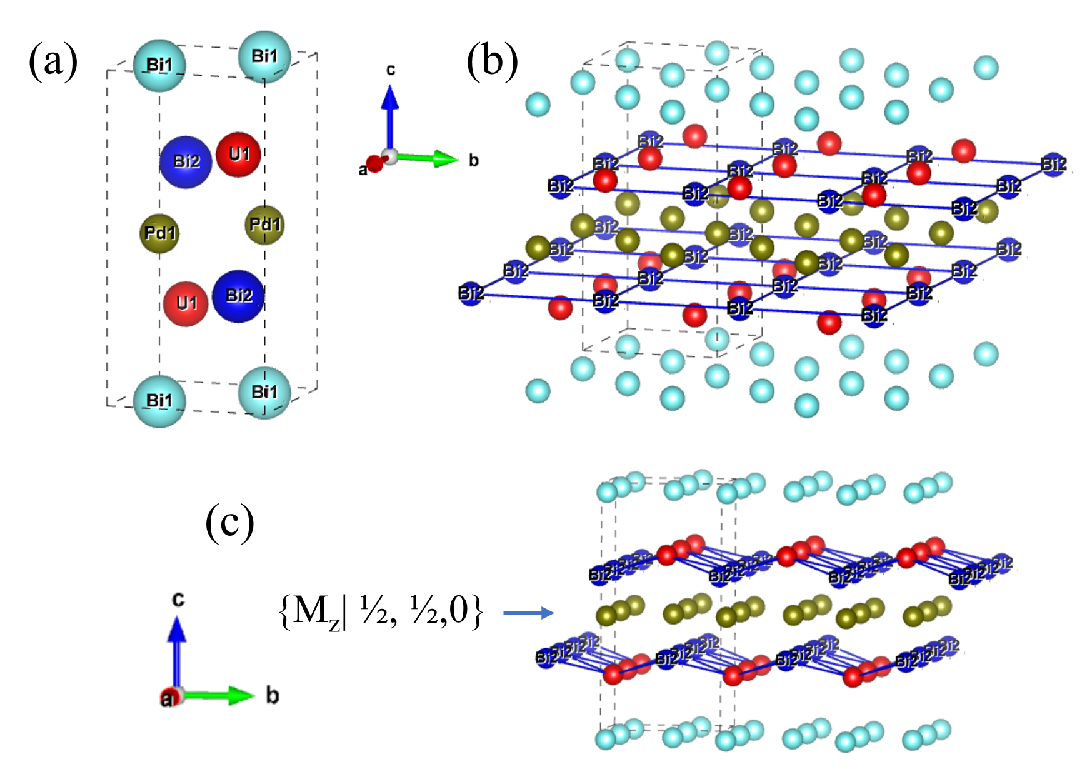}
\caption{\label{fig:UPdBi2structure}(a) HfCuSi$_2$-type tetragonal crystal structure of UPd$_{0.65}$Bi$_2$. The elements U, Pd, Bi(1), and Bi(2) are depicted in red, jade green, cyan and blue, respectively. The tetragonal unit cell is depicted by the dashed box. (b) shows the layers of square-nets of Bi(2) sites (in solid blue lines) stacked along the $c$ axis. (c) depicts the glide mirror plane $M_z$ in the crystal structure.}
\end{figure}

X-ray diffraction analysis confirms that UPd$_{0.65}$Bi$_2$ crystallizes in the HfCuSi$_2$-type tetragonal crystal structure within space group $P4/nmm$ (No. 129). Figure~\ref{fig:UPdBi2structure}(a) shows the crystal structure of UPd$_{0.65}$Bi$_2$, which displays two inequivalent Bismuth sites. Bi(2) sites occupy the Wyckoff position $2c$ (0.25 0.25 0.66751(11)) and form square nets stacked along the $c$ direction as shown in Figure \ref{fig:UPdBi2structure}(b). As discussed before, the $P4/nmm$ structure displays a nonsymmorphic symmetry operation: a glide mirror $ \{ M_z | \frac{1}{2}, \frac{1}{2}, 0 \}$ that corresponds to a reflection about a plane perpendicular to $c$ axis at $z=\frac{1}{2}$ and a translation along the $ab$ diagonal by ${\frac{1}{2}, \frac{1}{2}}$, as shown in Figure \ref{fig:UPdBi2structure}(c).

{\renewcommand{\arraystretch}{1.2}
\begin{table}[]
\caption{Single crystal refinement parameters for UPd$_{0.65}$Bi$_2$.}
\label{tab:refinement_parameters}

\begin{tabular}{ll}
\hhline{==}
Formula & UPd$_0.65$Bi$_2$
\\ \hline
Space Group  & P4/nmm
 \\
 Crystal Size (mm)  & $0.167 \times 0.113 \times 0.075$ \\
a ($\mathrm{\AA}$)                                                                           & 4.5478(6)  \\
c ($\mathrm{\AA}$)                                                                           & 9.320(2)  \\
V ($\mathrm{\AA}^{3}$)                        & 192.76(7)  \\
Z                                                                                            & 1          \\
T (K)                                                                                        & 293        \\
$\theta$ ($ \degree $)                                                                                     & 4.37 - 30.54 \\
$\mu$ (mm$^{-1}$)                                                                            & 618.614      \\
Measured/Independent Reflections                                                            & 7349/215       \\
R$_{int}$                                                                                    &  0.083          \\
Extinction Coefficient                                                                       & 0.0030(6)  \\
Refinement Parameters                                                                      & 13        \\
$\Delta \rho_{min}  - \Delta \rho_{max} (e \mathrm{\AA}^{-3}$) & -.3050 - 2.579      \\
$R (F > 2\sigma(F)$)\footnotemark       & 0.0253      \\
$wR( F^2)$\footnotemark & 0.0721      \\ 
GooF\footnotemark & 1.374 \\
\hhline{==}
\end{tabular}
\footnotetext[1]{$R(F) = \Sigma ||F_0|-|F_c||/\Sigma |F_0|$}
\footnotetext[2]{$ wR(F^2) = \bigl[\Sigma w(F_0^2 - F_c^2)^2/ \Sigma(F_0^2) \bigr]^{1/2} $}
\footnotetext[3]{ $GooF = \bigl( \bigl[ \Sigma w (F_0^2F_c^2)^2 \bigr]/(N_{\mathrm{ref}}- N_{\mathrm{param}}\bigr)^{1/2}$ for all reflections.}
\end{table}
}

Elemental analysis indicates the presence of significant palladium vacancies, of the order of 20$\%$. In fact, our single crystal x-ray refinement residual improves when the palladium occupancy is allowed to vary and yields UPd$_{0.65}$Bi$_2$. The final refined parameters are shown in Table~\ref{tab:refinement_parameters}. Vacancy at the transition metal site has been reported in other related compounds in this family \cite{tkachuk2004cerium, Sharma2023PRB,Rosa2015PRB}. The absence of superstructure modulations implies that the vacancies are not ordered, in contrast to other square-net materials \cite{Piva2021ACS, Lei, Patschke, DiMasi}, and points to robust square nets in UPd$_{0.65}$Bi$_2$. Atomic coordinates, isotropic displacement parameters, and occupancies are shown in Table~\ref{tab:atomic_sites} in Appendix 1.
\begin{figure}
	\centering
    \includegraphics[width=\linewidth]{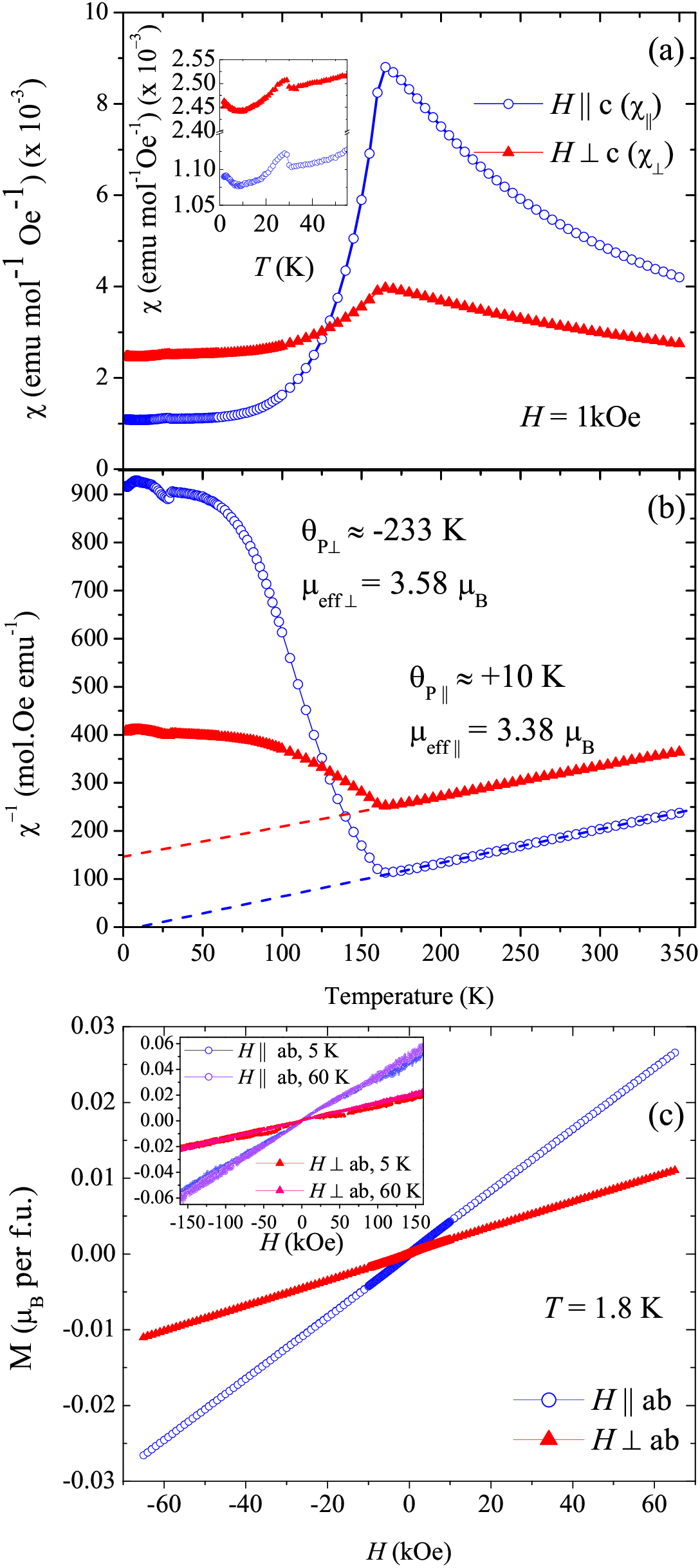}
\caption{\label{fig:ChivsTnew1} Magnetic properties of UPd$_{0.65}$Bi$_2$. Temperature dependence of (a) magnetic susceptibility and (b) inverse magnetic susceptibility in a single crystal of UPd$_{0.65}$Bi$_2$ for a magnetic field of 1~kOe applied parallel and perpendicular to the $c$ axis. Inset in (a) shows a zoomed in view of the transition at $T_1$ discussed in the main text. Dashed lines in (b) are Curie-Weiss fits to $\chi^{-1}(T)$ as described in the text. 
(c) Magnetic field dependence of magnetization for a magnetic field applied parallel and perpendicular to the $c$ axis. Inset shows magnetization at higher fields to 160 kOe. 
}
\end{figure}


\subsubsection{Susceptibility and Magnetization}

Figure \ref{fig:ChivsTnew1}(a) shows the temperature-dependent magnetic susceptibilities, $\chi_\parallel (T)$ and $\chi_\perp (T)$, 
 of a UPd$_{0.65}$Bi$_2$ single crystal for a small magnetic field  $H = 1$~kOe applied parallel ($H\parallel c$) and perpendicular ($H \perp c$) to the tetragonal $c$ axis, respectively. It is evident from the sharp cusp observed in $\chi (T)$ that UPd$_{0.65}$Bi$_2$ undergoes an antiferromagnetic transition at $T_{\text{N}} \simeq$164 K. Below $T_{\text{N}}$, a  reduction in magnetic susceptibility is observed for both $\chi_\parallel (T)$ and $\chi_\perp (T)$. The much larger reduction in $\chi_\parallel (T)$ suggests that the $[001]$ is the easy magnetization direction. Indeed, the magnetic structure of many members of this family was previously shown to be A-type with uranium moments aligned along the $c$ axis. In this structure, ferromagnetically aligned sheets of U moments in the $ab$ plane are stacked antiferromagnetically along the $c$ axis \cite{Kaczorowski}. However, in such a simple two sublattice antiferromagnetic structure, the antiparallel arrangement of moments along the high-symmetry principal axis (tetragonal $c$ axis) implies that $\chi_\perp (T)$ should remain mostly temperature-independent below $T_N$. The observed reduction in $\chi_\perp (T)$ therefore provides an important insight into the magnetic structure of UPd$_{0.65}$Bi$_2$: though the arrangement of magnetic moments in UPd$_{0.65}$Bi$_2$ is largely along the tetragonal $c$ axis and antiparallel, there is a small in-plane component due to canting or an incommensurate component is present.

 Upon further cooling, another feature is observed in both $\chi_\parallel (T)$ and $\chi_\perp (T)$  at $T_1 \simeq $ 30~K, as shown in the inset of Figure \ref{fig:ChivsTnew1}(a). This additional feature in $\chi (T)$ is likely caused by another magnetic transition due to a small change in the AFM magnetic structure. Previously, a similar behaviour was observed in UCuBi$_2$ \cite{KACZOROWSKI1992333} and USnTe \cite{TROC198767} and attributed to a spin reorientation. We note that the AFM transition temperature in UPd$_{0.65}$Bi$_2$,  $T_N \simeq$ 164~K, is only slightly lower than that of UNiBi$_2$ ($T_N \simeq$ 166~K), but a second transition is observed at lower temperature only in UPd$_{0.65}$Bi$_2$ and UCuBi$_2$, but not in UNiBi$_2$. Structurally, the $c$ axis in UPd$_{0.65}$Bi$_2$ ($\sim9.31$~\AA) is larger than that in UNiBi$_2$ ($\sim9.073$~\AA), but comparable to UCuBi$_2$ ($\sim9.376$~\AA). Though many parameters (e.g., the chemistry of the $d$-electron layer, unit cell volume) may be relevant to understand the magnetic trends in this family, we hypothesize that the $c$ axis parameter plays a major role in destabilizing the A-type magnetic structure and inducing a second phase transition at lower temperatures in UPd$_{0.65}$Bi$_2$ and UCuBi$_2$. 
 Indeed, in UCuBi$_2$, the magnetic susceptibility is qualitatively similar to $\chi(T)$ in UPd$_{0.65}$Bi$_2$ but with lower transition temperatures $T_{\text{N}}~=$~51~K and $T_1 =$~15~K \cite{Kaczorowski}. UAgBi$_2$, which has the largest $c$-axis parameter in this class of materials ($\sim10.3$~\AA), displays seven different magnetic phases as a function of temperature and magnetic field, which points to multiple competing interactions \cite{Gabriel2024}.

 The inverse susceptibilities, $\chi ^{-1}_\parallel(T)$ and $\chi ^{-1}_\perp(T)$, are shown in Figure \ref{fig:ChivsTnew1}(b). In the paramagnetic state, $\chi^{-1}(T)$ is well described by a Curie-Weiss (CW) expression, $\chi = {\text{C}}/(T-\theta)$, where $C$ is the Curie constant and $\theta$ is the Weiss temperature. It is worth noting that the magnetic susceptibility follows Curie-Weiss behavior all the way down to T$_N$. The effective magnetic moments obtained from CW fits, $\mu_{\text{eff}\perp}= $ 3.58$~\mu_B$ and $\mu_{\text{eff} \parallel}= $ 3.38$~\mu_B$, are very close to the effective magnetic moments $\mu_{\text{eff}} = 3.58~\mu_B$ and $\mu_{\text{eff}} = 3.62~\mu_B$ predicted by Hund's rule for U$^{4+}$ and U$^{3+}$ free ions, respectively, and indicates that the oxidation state of uranium cannot be determined by CW fits alone \cite{BUKOWSKI20043934}. The extracted Weiss temperatures are $\theta_{\parallel} =$+10 K and $\theta_{\perp} =$-233 K. In the simplest approximation using a molecular field in the absence of crystalline electric field (CEF) effects, the positive sign of $\theta$ would imply ferromagnetic interactions along the $c$ axis, whereas antiferromagnetic interactions would be expected in the $ab$ plane. The anisotropy of these interactions, however, is at odds with the A-type magnetism typically observed in this class of materials, wherein ferromagnetic layers in the $ab$ plane align antiferromagnetically along the $c$ axis. To determine whether this discrepancy is due to CEF effects, we attempted to fit $\chi (T)$ data to a point-charge CEF model plus mean-field anisotropic exchange terms. It is possible to reproduce the Weiss temperatures $\theta$ from CEF effects alone. However, due to the limited temperature range for the fits and the absence of clear CEF features in susceptibility, the fits are underconstrained and cannot provide reliable information on the CEF level splitting and wavefunctions.
 


We now turn to the field-dependent magnetization, $M($H$)$, of UPd$_{0.65}$Bi$_2$ at 1.8 K, shown in Figure \ref{fig:ChivsTnew1}(c). The main panel shows that the low-field magnetization curves are linear for both field directions, which is generally expected for AFM materials; however, the induced moments are very small. At 65~kOe, the induced moments are only $\mu_\parallel \approx 0.01 \mu_B$ and $\mu_\perp \approx 0.025 \mu_B$. In addition, $M($H$)$ data do not show either a field-induced transition or saturation to 160~kOe, as shown in the inset of Figure~\ref{fig:ChivsTnew1}(c). Our findings therefore indicate that the AFM structure is locked into a very stable configuration. At first sight, this behavior is unexpected given that most 112 materials display one or more field-induced transitions in this field range due to the competition between FM and AFM exchange interactions, the presence of higher-order spin exchange interactions, low-lying CEF levels, or local Dzyaloshinskii-Moriya (DM) interactions \cite{Simeth2024, Hayami, Seo}. However, the combination of a well-isolated CEF ground state and a dominant AFM exchange interaction that is much larger than competing interactions can describe our observations.   \\



\subsubsection{Heat Capacity}
\begin{figure}[!hb]
	\centering
    \includegraphics[width=\columnwidth]{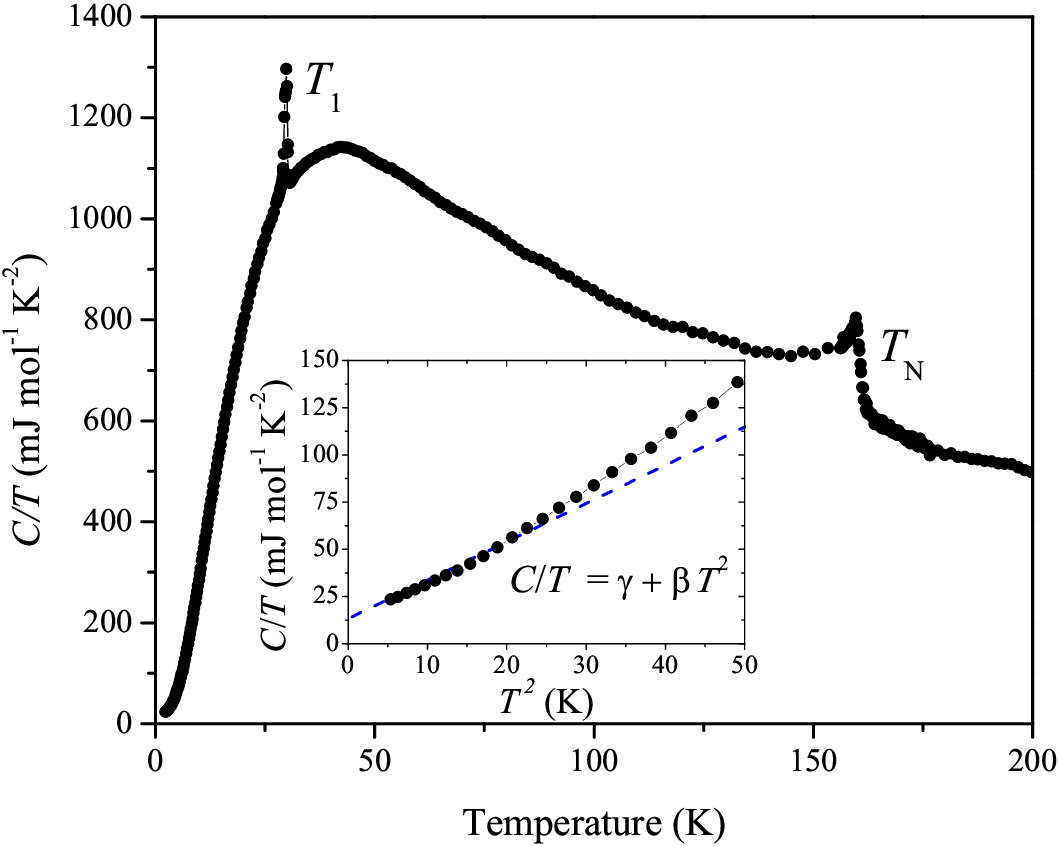}
\caption{\label{fig:CbyTvsT} Temperature dependance of heat capacity $C/T$ of UPd$_{0.65}$Bi$_2$. The antiferromagnetic transition at $T_N$ and the first order transition at $T_1$ are labelled. Inset shows a linear fit to $C/T$ vs $T^2$ extracting  $\gamma$ as the y-intercept. }
\end{figure}

The temperature-dependent specific heat, $C/T$, of a UPd$_{0.65}$Bi$_2$ single crystal is shown in Fig. \ref{fig:CbyTvsT}. $C/T$ exhibits two distinct features at $T_{\text{N}} \simeq 161$~K and $T_1 \simeq$~30 K.
The feature at  $T_{\text{N}}$ has a prototypical lambda-like shape of a second-order phase transition and corresponds to the paramagnetic to antiferromagnetic phase transition in UPd$_{0.65}$Bi$_2$. Further, this transition is noticeably mean-field like suggesting negligible fluctuations above the transition. In contrast, the feature at $T_1$ is sharp and largely symmetric, which indicates a first-ordered phase transition and likely corresponds to a change of magnetic structure in agreement with magnetic susceptibility. 

\begin{figure*}[!ht]
	\centering
    \includegraphics[width=\linewidth]{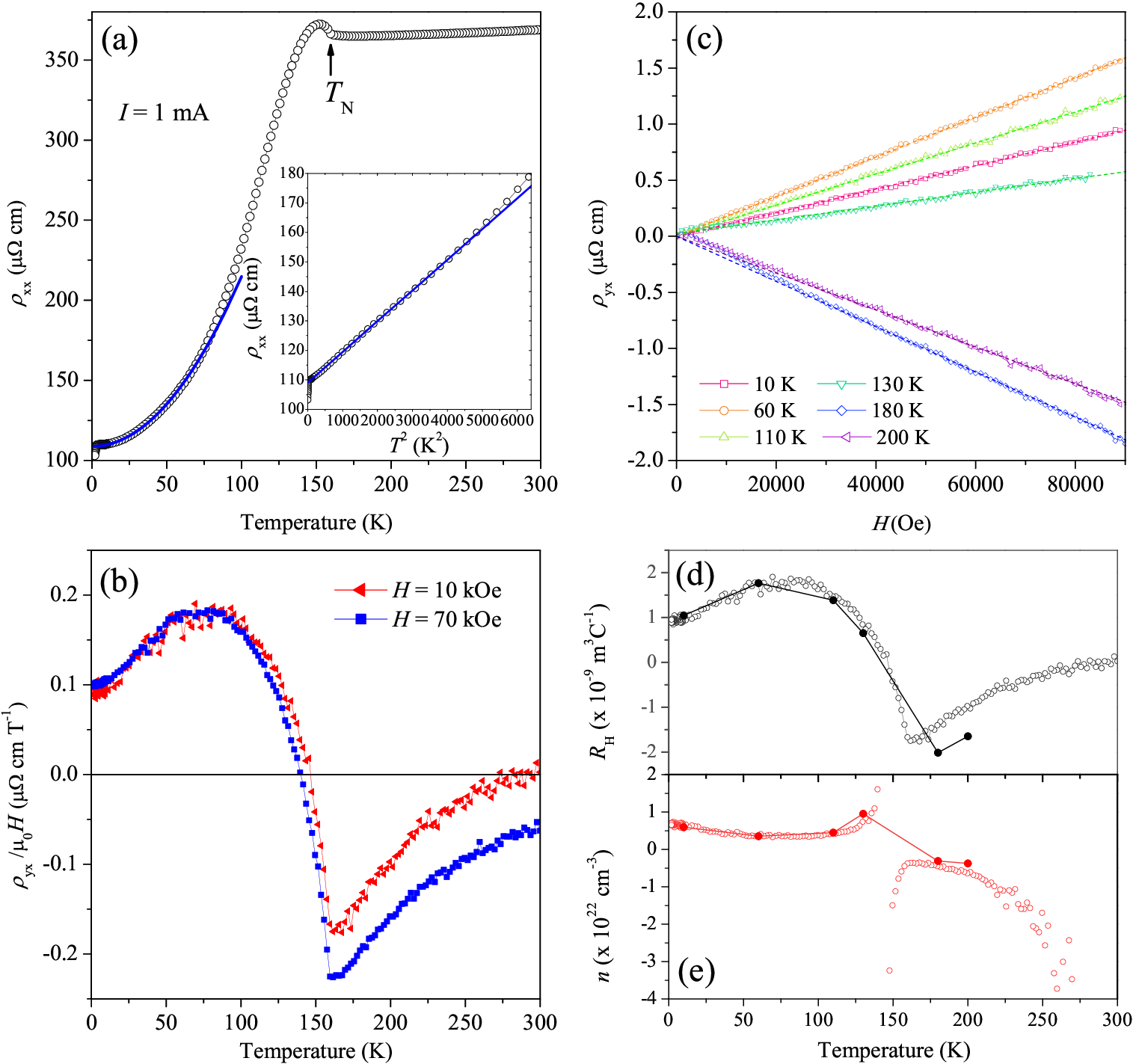}
\caption{\label{fig:RvsT} Electrical transport properties of UPd$_{0.65}$Bi$_2$. Temperature dependance of in-plane resistivity in a UPd$_{0.65}$Bi$_2$ single crystal for a current $I =$ 1 mA. The solid lines are fits to the resistivity as discussed in the text. Inset shows a Fermi-liquid term to the low-temperature resistivity as discussed in text. 
(b) Temperature- and (c) magnetic field-dependent Hall resistivity $\rho_{yx}$ for $I = 1$mA and field along the $c$ axis. (d) The Hall coefficient $R_H$ and (e) carrier density $n$ obtained from the temperature- (open symbols) and field- (solid symbols) dependent Hall resistance.}
\end{figure*}

Unfortunately, a non-$f$ analog compound of UPd$_{0.65}$Bi$_2$ could not be grown to subtract the non-magnetic contributions to heat capacity. Therefore, the electronic contribution is deduced by fitting the low-temperature heat capacity with $C/T = \gamma +\beta T^2$, where $\gamma$ is the electronic contribution or the Sommerfield coefficient and $\beta$ is the bosonic contribution. A small $\gamma = $ ~13 mJmol$^{-1}$K$^{-2}$ suggests weak Kondo coupling and that the $5f$ electrons are predominantly localized.\\

\subsubsection{Electrical Resistivity}

The temperature-dependent in-plane resistivity, $\rho_{ab}(T)$, is shown in Figure \ref{fig:RvsT}(a). On cooling, $\rho_{ab}(T)$ decreases monotonically down to the antiferromagnetic transition at T$_N$. The room-temperature resistivity $\rho_{300\text{K}} \simeq$ 370 $\mu \Omega \text{cm}$ is comparable to other uranium antimonides \cite{Kaczorowski} and bismuthides \cite{PhysRevB.92.104425}. The electrical transport in UPd$_{0.65}$Bi$_2$ is consistent with a semimetallic behavior similar to other 112 uranium based materials \cite{Kaczorowski}. At $T_{\text{N}} \simeq $161~K, an increase in $\rho_{ab}(T)$ is observed and points to the opening of a gap in the electronic structure. In metallic systems, the electrical resistivity at a second-order magnetic transition is expected to follow the Fisher-Langer behavior wherein d$\rho$/dT is proportional to the heat capacity with a positive constant of proportionality \cite{Fisher, Zumsteg, Yoon}.
The resistivity in ferromagnets or commensurate antiferromagnets therefore decreases as the temperature is lowered through the transition. However, an upward inflection may occur in incommensurate antiferromagnets due to the opening of the so-called superzone gap \cite{Mackintosh, Elliott} when the periodicity of the magnetic structure is different from the periodicity of the lattice \cite{Becker, Onimaru, Yejun, Ru}.


In the magnetically ordered state, $\rho_{ab}(T)$ decreases rapidly on cooling. 
Below 75~K, the resistivity can be well described by a Fermi-liquid (FL) term, $\rho(T) = \rho_0 + AT^2$, with $\rho_0 = 110 ~\mu\Omega \text{cm}$ and $A = 0.01~\mu\Omega\text{cmK}^{-2}$ (solid line in the inset of Figure~\ref{fig:RvsT}). We note that the abrupt drop in $\rho_{ab}(T)$ observed around 3~K corresponds to superconducting BiPd binaries usually present as impurity filaments in single crystals grown in excess Bismuth flux. Previously, we observed such superconducting impurity binaries in Ce$_3$Bi$_4$Pd$_3$ \cite{Ajeesh} and CeRh$_2$As$_2$ \cite{Mishra} single crystals grown in Bismuth flux. By combining the FL coefficient and the Sommerfeld coefficient, we extract the Kadowaki-Woods (KW) ratio $\frac{A}{\gamma^2} \approx 5.9\times 10^{-5}\mu\Omega\text{cm} (\frac{\text{mol K}}{\text{mJ}})^2$, which is approximately 6 times larger than that observed in other heavy-fermion compounds \cite{Kadowaki}. This larger KW ratio in UPd$_{0.65}$Bi$_2$ cannot be due to an $N>2$ ground state degeneracy (where $N = 2J+1$) because $N>2$ leads to a smaller KW ratio \cite{Tsujii}. A recent generalization of the KW ratio in strongly correlated 
materials \cite{Jacko} provides a modified universal ratio $A/\gamma^{2}=C/f_{dx}(n)$, where $C$ is a constant, $f_{dx}(n)\equiv n D_{0}^{2} \langle v^{2}_{0x}\rangle\xi^{2}$, $n$ is the carrier density, $D_{0}$ is the bare density of states at the Fermi level, $\langle v^{2}_{0x}\rangle$ is the unrenormalized velocity averaged over the Fermi surface, and $\xi \approx 1$. Importantly, this ratio is materials dependent and will vary depending on the carrier density, velocity, and dimensionality of the system.
Because 112 uranium-based materials are known to be semimetals, a sensible explanation for the larger KW ratio in UPd$_{0.65}$Bi$_2$ is a smaller bare density of states compared to heavy-fermion metals. Band structure calculations or experiments would be valuable to confirm this scenario.


Figure \ref{fig:RvsT}(b) shows the temperature-dependent Hall resistivity, $\rho_{\text{yx}}(T)$, with current in the $ab$ plane and magnetic fields of 10~kOe and 70~kOe applied along the $c$ axis. The Hall resistivity is obtained by antisymmetrizing the Hall data with respect to positive $(H_+)$ and negative $(H_-)$ field as $\rho_{yx}=[\rho_{yx}(H_+)-\rho_{yx}(H_-)]/2$. As evident in Figure \ref{fig:RvsT}(b), $\rho_{yx}$ is much smaller ($< 1\%$) than the longitudinal resistivity, $\rho_{xx}$, and its negative sign at room temperature indicates the primary carriers in the paramagnetic state are electrons. Upon cooling, $\rho_{yx}(T)$ increases in the negative direction before an abrupt change at the ordering temperature. At $T_{\text{N}}$, $\rho_{yx}(T)$ sharply becomes positive, which indicates hole-type carriers dominate electronic transport in the ordered state.

Magnetic field dependence of the Hall resistivity $\rho_{yx}$($H$) is shown in Figure \ref{fig:RvsT}(c) at several different temperatures across $T_N$ and $T_1$. $\rho_{yx}$($H$) is linear at $H> 5$ kOe and a single-band model can be used to describe the transport properties. The origin of the small non-linearity at very low fields below $T_{\text{N}}$ is presently unclear. 
Below 80~K, however, $\rho_{yx}$($H$) becomes linear down to zero field. We therefore obtain the Hall coefficient from the slope of $\rho_{yx}$($H$) as $R_H = \rho_{yx}/\mu_0 H$ and the corresponding carrier density as $n = 1/eR_H$. The calculations of the Hall coefficient, $R_H$, and the carrier density, $n$, for UPd$_{0.65}$Bi$_2$ are shown in Figures~\ref{fig:RvsT}(d) and (e), respectively. The increase in $\rho_{yx}$ (i.e., the decrease in carrier density) on cooling is also consistent with semimetallic behavior in UPd$_{0.65}$Bi$_2$. Further, it is evident that a drastic change in $R_H$, and consequently $n$, occurs across the antiferromagnetic transition. The abrupt change in $\rho_{yx}(T)$, accompanied by a gradual change of sign across the paramagnetic-antiferromagnetic transition, suggests that the electron pockets are gapped in the ordered-state and hole-Fermi surfaces become the dominant carriers in the antiferromagnetic state. 

\subsection{Summary}
In summary, we synthesized UPd$_{0.65}$Bi$_2$, a new member of the U$MX_2$ family $(M =$ transition metal, $X =$ pnictogen), and investigated its structural, electronic, magnetic and thermodynamic properties. Our magnetic susceptibility, magnetization, heat capacity and resistivity measurements reveal that UPd$_{0.65}$Bi$_2$ orders antiferromagnetically below $T_N \approx$ 161~K and undergoes another magnetic phase transition at $T_1 \approx 30~$K likely related to a change in the antiferromagnetic structure. Electrical resistivity data indicates the opening of a superzone gap at the Fermi surface consistent with an an incommensurate magnetic structure, whereas Hall effect reveals a drastic sign change in the carrier density across $T_N$. A small $\gamma \simeq 13~ \text{mJmol}^{-1}\text{K}^{-2}$ suggests that UPd$_{0.65}$Bi$_2$ falls in the localized $f$-electron limit. Therefore, the nonsymmorphic structure of UPd$_{0.65}$Bi$_2$ is not expected to give rise to a Weyl-Kondo semi-metallic state. In contrast, our electronic and magnetic measurements provide evidence for an incommensurate magnetic structure that goes beyond the conventional A-type (++- -) magnetism expected in this class of materials. Neutron diffraction and magnetization measurements at higher fields would be valuable to elucidate the presence of unconventional spin textures.

\begin{acknowledgments}
Work at Los Alamos National Laboratory was performed under the auspices of the U.S. Department of Energy, Office of Basic Energy Sciences, Division of Materials Science and Engineering. S. M. and C. S. K. acknowledge support from the Laboratory Directed Research and Development program. Scanning electron microscope and energy dispersive x-ray measurements were performed at the Electron Microscopy Lab and supported by the Center for Integrated Nanotechnologies, an Office of Science User Facility operated for the U.S. Department of Energy Office of Science.
\end{acknowledgments}

\subsection{Appendix 1. Single crystal x-ray diffraction analysis}

Table~\ref{tab:atomic_sites} shows the atomic coordinates, isotropic displacement parameters, and occupancies of UPd$_{0.65}$Bi$_2$.

{\renewcommand{\arraystretch}{1.5}
\begin{table*}[]
\caption{Fractional atomic coordinates and isotropic displacement parameters of the refinement of UPd$_{0.65}$Bi$_2$. Off-diagonal elements are 0 by symmetry.}
\label{tab:atomic_sites}
\begin{tabularx}{\textwidth}{XXXXXXXXXX}

\hhline{==========}

 Atom & Wyckoff & x   & y   & z   & Occupancy  & U$_{\mathrm{iso}}$ ($\mathrm{\AA}^2$)   &U$_{11}$ ($\mathrm{\AA}^2$)   & U$_{22}$  ($\mathrm{\AA}^2$)  & U$_{33}$  ($\mathrm{\AA}^2$)  \\ \hline
 U & 2c & 1/4 & 1/4 & 0.27085(8)  & 1 & 0.0141(3) & 0.0147(3) &  0.0147(3) & 0.0130(4)\\
 Pd & 2b & 3/4 & 1/4 & 1/2 & 0.651(9)& 0.0177(8) & 0.0202(10) & 0.0202(10) & 0.0127(13) \\
 Bi2 & 2c & 1/4 & 1/4 & 0.66751(11)  & 1 & 0.0216(3) &  0.0183(3) & 0.0183(3) & 0.0280(5)\\
 Bi1 & 2a & 3/4 & 1/4 & 0 & 1 & 0.0150(3) & 0.0154(3) & 0.0154(3) & 0.0142(4)  \\
 \hline

\hhline{==========}
\end{tabularx}
\end{table*}
}

\bibliography{UPdBi2_V2}

\begin{thebibliography}{51}%
\makeatletter
\providecommand \@ifxundefined [1]{%
 \@ifx{#1\undefined}
}%
\providecommand \@ifnum [1]{%
 \ifnum #1\expandafter \@firstoftwo
 \else \expandafter \@secondoftwo
 \fi
}%
\providecommand \@ifx [1]{%
 \ifx #1\expandafter \@firstoftwo
 \else \expandafter \@secondoftwo
 \fi
}%
\providecommand \natexlab [1]{#1}%
\providecommand \enquote  [1]{``#1''}%
\providecommand \bibnamefont  [1]{#1}%
\providecommand \bibfnamefont [1]{#1}%
\providecommand \citenamefont [1]{#1}%
\providecommand \href@noop [0]{\@secondoftwo}%
\providecommand \href [0]{\begingroup \@sanitize@url \@href}%
\providecommand \@href[1]{\@@startlink{#1}\@@href}%
\providecommand \@@href[1]{\endgroup#1\@@endlink}%
\providecommand \@sanitize@url [0]{\catcode `\\12\catcode `\$12\catcode
  `\&12\catcode `\#12\catcode `\^12\catcode `\_12\catcode `\%12\relax}%
\providecommand \@@startlink[1]{}%
\providecommand \@@endlink[0]{}%
\providecommand \url  [0]{\begingroup\@sanitize@url \@url }%
\providecommand \@url [1]{\endgroup\@href {#1}{\urlprefix }}%
\providecommand \urlprefix  [0]{URL }%
\providecommand \Eprint [0]{\href }%
\providecommand \doibase [0]{https://doi.org/}%
\providecommand \selectlanguage [0]{\@gobble}%
\providecommand \bibinfo  [0]{\@secondoftwo}%
\providecommand \bibfield  [0]{\@secondoftwo}%
\providecommand \translation [1]{[#1]}%
\providecommand \BibitemOpen [0]{}%
\providecommand \bibitemStop [0]{}%
\providecommand \bibitemNoStop [0]{.\EOS\space}%
\providecommand \EOS [0]{\spacefactor3000\relax}%
\providecommand \BibitemShut  [1]{\csname bibitem#1\endcsname}%
\let\auto@bib@innerbib\@empty
\bibitem [{\citenamefont {Qi}\ and\ \citenamefont {Zhang}(2011)}]{Qi}%
  \BibitemOpen
  \bibfield  {author} {\bibinfo {author} {\bibfnamefont {X.-L.}\ \bibnamefont
  {Qi}}\ and\ \bibinfo {author} {\bibfnamefont {S.-C.}\ \bibnamefont {Zhang}},\
  }\bibfield  {title} {\bibinfo {title} {Topological insulators and
  superconductors},\ }\href {https://doi.org/10.1103/RevModPhys.83.1057}
  {\bibfield  {journal} {\bibinfo  {journal} {Rev. Mod. Phys.}\ }\textbf
  {\bibinfo {volume} {83}},\ \bibinfo {pages} {1057} (\bibinfo {year}
  {2011})}\BibitemShut {NoStop}%
\bibitem [{\citenamefont {Lian}\ \emph {et~al.}(2018)\citenamefont {Lian},
  \citenamefont {Sun}, \citenamefont {Vaezi}, \citenamefont {Qi},\ and\
  \citenamefont {Zhang}}]{Biao}%
  \BibitemOpen
  \bibfield  {author} {\bibinfo {author} {\bibfnamefont {B.}~\bibnamefont
  {Lian}}, \bibinfo {author} {\bibfnamefont {X.-Q.}\ \bibnamefont {Sun}},
  \bibinfo {author} {\bibfnamefont {A.}~\bibnamefont {Vaezi}}, \bibinfo
  {author} {\bibfnamefont {X.-L.}\ \bibnamefont {Qi}},\ and\ \bibinfo {author}
  {\bibfnamefont {S.-C.}\ \bibnamefont {Zhang}},\ }\bibfield  {title} {\bibinfo
  {title} {Topological quantum computation based on chiral majorana fermions},\
  }\href {https://doi.org/10.1073/pnas.1810003115} {\bibfield  {journal}
  {\bibinfo  {journal} {Proceedings of the National Academy of Sciences}\
  }\textbf {\bibinfo {volume} {115}},\ \bibinfo {pages} {10938} (\bibinfo
  {year} {2018})}\BibitemShut {NoStop}%
\bibitem [{\citenamefont {Gilbert}(2021)}]{Gilbert}%
  \BibitemOpen
  \bibfield  {author} {\bibinfo {author} {\bibfnamefont {M.~J.}\ \bibnamefont
  {Gilbert}},\ }\bibfield  {title} {\bibinfo {title} {Topological
  electronics},\ }\href {https://doi.org/10.1038/s42005-021-00569-5} {\bibfield
   {journal} {\bibinfo  {journal} {Commun Phys}\ }\textbf {\bibinfo {volume}
  {4}},\ \bibinfo {pages} {70} (\bibinfo {year} {2021})}\BibitemShut {NoStop}%
\bibitem [{\citenamefont {He}\ \emph {et~al.}(2021)\citenamefont {He},
  \citenamefont {Hughes}, \citenamefont {Armitage}, \citenamefont {Tokura},\
  and\ \citenamefont {Wang}}]{He}%
  \BibitemOpen
  \bibfield  {author} {\bibinfo {author} {\bibfnamefont {Q.~L.}\ \bibnamefont
  {He}}, \bibinfo {author} {\bibfnamefont {T.~L.}\ \bibnamefont {Hughes}},
  \bibinfo {author} {\bibfnamefont {N.~P.}\ \bibnamefont {Armitage}}, \bibinfo
  {author} {\bibfnamefont {Y.}~\bibnamefont {Tokura}},\ and\ \bibinfo {author}
  {\bibfnamefont {K.~L.}\ \bibnamefont {Wang}},\ }\bibfield  {title} {\bibinfo
  {title} {Topological spintronics and magnetoelectronics},\ }\href
  {https://doi.org/10.1038/s41563-021-01138-5} {\bibfield  {journal} {\bibinfo
  {journal} {Nat. Mater.}\ }\textbf {\bibinfo {volume} {21}},\ \bibinfo {pages}
  {15} (\bibinfo {year} {2021})}\BibitemShut {NoStop}%
\bibitem [{\citenamefont {\ifmmode~\check{S}\else \v{S}\fi{}mejkal}\ \emph
  {et~al.}(2018)\citenamefont {\ifmmode~\check{S}\else \v{S}\fi{}mejkal},
  \citenamefont {Mokrousov}, \citenamefont {Yan},\ and\ \citenamefont
  {MacDonald}}]{Smejkal}%
  \BibitemOpen
  \bibfield  {author} {\bibinfo {author} {\bibfnamefont {L.}~\bibnamefont
  {\ifmmode~\check{S}\else \v{S}\fi{}mejkal}}, \bibinfo {author} {\bibfnamefont
  {Y.}~\bibnamefont {Mokrousov}}, \bibinfo {author} {\bibfnamefont
  {B.}~\bibnamefont {Yan}},\ and\ \bibinfo {author} {\bibfnamefont {A.~H.}\
  \bibnamefont {MacDonald}},\ }\bibfield  {title} {\bibinfo {title}
  {Topological antiferromagnetic spintronics},\ }\href
  {https://doi.org/10.1038/s41567-018-0064-5} {\bibfield  {journal} {\bibinfo
  {journal} {Nature Phys}\ }\textbf {\bibinfo {volume} {14}},\ \bibinfo {pages}
  {242–251} (\bibinfo {year} {2018})}\BibitemShut {NoStop}%
\bibitem [{\citenamefont {Schoop}\ \emph {et~al.}(2016)\citenamefont {Schoop},
  \citenamefont {Ali}, \citenamefont {Straßer}, \citenamefont {Topp},
  \citenamefont {Varykhalov}, \citenamefont {Marchenko}, \citenamefont
  {Duppel}, \citenamefont {Parkin}, \citenamefont {Lotsch},\ and\ \citenamefont
  {Ast}}]{Schoop}%
  \BibitemOpen
  \bibfield  {author} {\bibinfo {author} {\bibfnamefont {L.~M.}\ \bibnamefont
  {Schoop}}, \bibinfo {author} {\bibfnamefont {M.~N.}\ \bibnamefont {Ali}},
  \bibinfo {author} {\bibfnamefont {C.}~\bibnamefont {Straßer}}, \bibinfo
  {author} {\bibfnamefont {A.}~\bibnamefont {Topp}}, \bibinfo {author}
  {\bibfnamefont {A.}~\bibnamefont {Varykhalov}}, \bibinfo {author}
  {\bibfnamefont {D.}~\bibnamefont {Marchenko}}, \bibinfo {author}
  {\bibfnamefont {V.}~\bibnamefont {Duppel}}, \bibinfo {author} {\bibfnamefont
  {S.~S.~P.}\ \bibnamefont {Parkin}}, \bibinfo {author} {\bibfnamefont {B.~V.}\
  \bibnamefont {Lotsch}},\ and\ \bibinfo {author} {\bibfnamefont {C.~R.}\
  \bibnamefont {Ast}},\ }\bibfield  {title} {\bibinfo {title} {Dirac cone
  protected by non-symmorphic symmetry and three-dimensional dirac line node in
  {Z}r{S}i{S}},\ }\href {https://doi.org/10.1038/ncomms11696} {\bibfield
  {journal} {\bibinfo  {journal} {Nature Commun}\ }\textbf {\bibinfo {volume}
  {7}},\ \bibinfo {pages} {11696} (\bibinfo {year} {2016})}\BibitemShut
  {NoStop}%
\bibitem [{\citenamefont {Schoop}\ \emph {et~al.}(2018)\citenamefont {Schoop},
  \citenamefont {Topp}, \citenamefont {Lippmann}, \citenamefont {Orlandi},
  \citenamefont {Müchler}, \citenamefont {Vergniory}, \citenamefont {Sun},
  \citenamefont {Rost}, \citenamefont {Duppel}, \citenamefont {Krivenkov},
  \citenamefont {Sheoran}, \citenamefont {Manuel}, \citenamefont {Varykhalov},
  \citenamefont {Yan}, \citenamefont {Kremer}, \citenamefont {Ast},\ and\
  \citenamefont {Lotsch}}]{Schoop2018}%
  \BibitemOpen
  \bibfield  {author} {\bibinfo {author} {\bibfnamefont {L.~M.}\ \bibnamefont
  {Schoop}}, \bibinfo {author} {\bibfnamefont {A.}~\bibnamefont {Topp}},
  \bibinfo {author} {\bibfnamefont {J.}~\bibnamefont {Lippmann}}, \bibinfo
  {author} {\bibfnamefont {F.}~\bibnamefont {Orlandi}}, \bibinfo {author}
  {\bibfnamefont {L.}~\bibnamefont {Müchler}}, \bibinfo {author}
  {\bibfnamefont {M.~G.}\ \bibnamefont {Vergniory}}, \bibinfo {author}
  {\bibfnamefont {Y.}~\bibnamefont {Sun}}, \bibinfo {author} {\bibfnamefont
  {A.~W.}\ \bibnamefont {Rost}}, \bibinfo {author} {\bibfnamefont
  {V.}~\bibnamefont {Duppel}}, \bibinfo {author} {\bibfnamefont
  {M.}~\bibnamefont {Krivenkov}}, \bibinfo {author} {\bibfnamefont
  {S.}~\bibnamefont {Sheoran}}, \bibinfo {author} {\bibfnamefont
  {P.}~\bibnamefont {Manuel}}, \bibinfo {author} {\bibfnamefont
  {A.}~\bibnamefont {Varykhalov}}, \bibinfo {author} {\bibfnamefont
  {B.}~\bibnamefont {Yan}}, \bibinfo {author} {\bibfnamefont {R.~K.}\
  \bibnamefont {Kremer}}, \bibinfo {author} {\bibfnamefont {C.~R.}\
  \bibnamefont {Ast}},\ and\ \bibinfo {author} {\bibfnamefont {B.~V.}\
  \bibnamefont {Lotsch}},\ }\bibfield  {title} {\bibinfo {title} {Tunable weyl
  and dirac states in the nonsymmorphic compound {C}e{S}b{T}e},\ }\href
  {https://doi.org/10.1126/sciadv.aar2317} {\bibfield  {journal} {\bibinfo
  {journal} {Science Advances}\ }\textbf {\bibinfo {volume} {4}},\ \bibinfo
  {pages} {eaar2317} (\bibinfo {year} {2018})}\BibitemShut {NoStop}%
\bibitem [{\citenamefont {Watanabe}\ \emph {et~al.}(2016)\citenamefont
  {Watanabe}, \citenamefont {Po}, \citenamefont {Zaletel},\ and\ \citenamefont
  {Vishwanath}}]{Watanabe}%
  \BibitemOpen
  \bibfield  {author} {\bibinfo {author} {\bibfnamefont {H.}~\bibnamefont
  {Watanabe}}, \bibinfo {author} {\bibfnamefont {H.~C.}\ \bibnamefont {Po}},
  \bibinfo {author} {\bibfnamefont {M.~P.}\ \bibnamefont {Zaletel}},\ and\
  \bibinfo {author} {\bibfnamefont {A.}~\bibnamefont {Vishwanath}},\ }\bibfield
   {title} {\bibinfo {title} {Filling-enforced gaplessness in band structures
  of the 230 space groups},\ }\href
  {https://doi.org/10.1103/PhysRevLett.117.096404} {\bibfield  {journal}
  {\bibinfo  {journal} {Phys. Rev. Lett.}\ }\textbf {\bibinfo {volume} {117}},\
  \bibinfo {pages} {096404} (\bibinfo {year} {2016})}\BibitemShut {NoStop}%
\bibitem [{\citenamefont {Po}\ \emph {et~al.}(2017)\citenamefont {Po},
  \citenamefont {Vishwanath},\ and\ \citenamefont {Watanabe}}]{Po}%
  \BibitemOpen
  \bibfield  {author} {\bibinfo {author} {\bibfnamefont {H.~C.}\ \bibnamefont
  {Po}}, \bibinfo {author} {\bibfnamefont {A.}~\bibnamefont {Vishwanath}},\
  and\ \bibinfo {author} {\bibfnamefont {H.}~\bibnamefont {Watanabe}},\
  }\bibfield  {title} {\bibinfo {title} {Symmetry-based indicators of band
  topology in the 230 space groups},\ }\href
  {https://doi.org/10.1038/s41467-017-00133-2} {\bibfield  {journal} {\bibinfo
  {journal} {Nature Commun}\ }\textbf {\bibinfo {volume} {8}},\ \bibinfo
  {pages} {50} (\bibinfo {year} {2017})}\BibitemShut {NoStop}%
\bibitem [{\citenamefont {Bian}\ \emph
  {et~al.}(2016{\natexlab{a}})\citenamefont {Bian}, \citenamefont {Chang},
  \citenamefont {Sankar}, \citenamefont {Xu}, \citenamefont {Zheng},
  \citenamefont {Neupert}, \citenamefont {Chiu}, \citenamefont {Huang},
  \citenamefont {Chang}, \citenamefont {Belopolski}, \citenamefont {Sanchez},
  \citenamefont {Neupane}, \citenamefont {Alidoust}, \citenamefont {Liu},
  \citenamefont {Wang}, \citenamefont {Lee}, \citenamefont {Jeng},
  \citenamefont {Zhang}, \citenamefont {Yuan}, \citenamefont {Jia},
  \citenamefont {Bansil}, \citenamefont {Chou}, \citenamefont {Lin},\ and\
  \citenamefont {Hasan}}]{Bian1}%
  \BibitemOpen
  \bibfield  {author} {\bibinfo {author} {\bibfnamefont {G.}~\bibnamefont
  {Bian}}, \bibinfo {author} {\bibfnamefont {T.-R.}\ \bibnamefont {Chang}},
  \bibinfo {author} {\bibfnamefont {R.}~\bibnamefont {Sankar}}, \bibinfo
  {author} {\bibfnamefont {S.-Y.}\ \bibnamefont {Xu}}, \bibinfo {author}
  {\bibfnamefont {H.}~\bibnamefont {Zheng}}, \bibinfo {author} {\bibfnamefont
  {T.}~\bibnamefont {Neupert}}, \bibinfo {author} {\bibfnamefont {C.-K.}\
  \bibnamefont {Chiu}}, \bibinfo {author} {\bibfnamefont {S.-M.}\ \bibnamefont
  {Huang}}, \bibinfo {author} {\bibfnamefont {G.}~\bibnamefont {Chang}},
  \bibinfo {author} {\bibfnamefont {I.}~\bibnamefont {Belopolski}}, \bibinfo
  {author} {\bibfnamefont {D.~S.}\ \bibnamefont {Sanchez}}, \bibinfo {author}
  {\bibfnamefont {M.}~\bibnamefont {Neupane}}, \bibinfo {author} {\bibfnamefont
  {N.}~\bibnamefont {Alidoust}}, \bibinfo {author} {\bibfnamefont
  {C.}~\bibnamefont {Liu}}, \bibinfo {author} {\bibfnamefont {B.}~\bibnamefont
  {Wang}}, \bibinfo {author} {\bibfnamefont {C.-C.}\ \bibnamefont {Lee}},
  \bibinfo {author} {\bibfnamefont {H.-T.}\ \bibnamefont {Jeng}}, \bibinfo
  {author} {\bibfnamefont {C.}~\bibnamefont {Zhang}}, \bibinfo {author}
  {\bibfnamefont {Z.}~\bibnamefont {Yuan}}, \bibinfo {author} {\bibfnamefont
  {S.}~\bibnamefont {Jia}}, \bibinfo {author} {\bibfnamefont {A.}~\bibnamefont
  {Bansil}}, \bibinfo {author} {\bibfnamefont {F.}~\bibnamefont {Chou}},
  \bibinfo {author} {\bibfnamefont {H.}~\bibnamefont {Lin}},\ and\ \bibinfo
  {author} {\bibfnamefont {M.~Z.}\ \bibnamefont {Hasan}},\ }\bibfield  {title}
  {\bibinfo {title} {Topological nodal-line fermions in spin-orbit metal
  {P}b{T}a{S}e$_2$},\ }\href {https://doi.org/10.1038/ncomms10556} {\bibfield
  {journal} {\bibinfo  {journal} {Nature Commun}\ }\textbf {\bibinfo {volume}
  {7}},\ \bibinfo {pages} {10556} (\bibinfo {year}
  {2016}{\natexlab{a}})}\BibitemShut {NoStop}%
\bibitem [{\citenamefont {Bian}\ \emph
  {et~al.}(2016{\natexlab{b}})\citenamefont {Bian}, \citenamefont {Chang},
  \citenamefont {Zheng}, \citenamefont {Velury}, \citenamefont {Xu},
  \citenamefont {Neupert}, \citenamefont {Chiu}, \citenamefont {Huang},
  \citenamefont {Sanchez}, \citenamefont {Belopolski}, \citenamefont
  {Alidoust}, \citenamefont {Chen}, \citenamefont {Chang}, \citenamefont
  {Bansil}, \citenamefont {Jeng}, \citenamefont {Lin},\ and\ \citenamefont
  {Hasan}}]{Bian}%
  \BibitemOpen
  \bibfield  {author} {\bibinfo {author} {\bibfnamefont {G.}~\bibnamefont
  {Bian}}, \bibinfo {author} {\bibfnamefont {T.-R.}\ \bibnamefont {Chang}},
  \bibinfo {author} {\bibfnamefont {H.}~\bibnamefont {Zheng}}, \bibinfo
  {author} {\bibfnamefont {S.}~\bibnamefont {Velury}}, \bibinfo {author}
  {\bibfnamefont {S.-Y.}\ \bibnamefont {Xu}}, \bibinfo {author} {\bibfnamefont
  {T.}~\bibnamefont {Neupert}}, \bibinfo {author} {\bibfnamefont {C.-K.}\
  \bibnamefont {Chiu}}, \bibinfo {author} {\bibfnamefont {S.-M.}\ \bibnamefont
  {Huang}}, \bibinfo {author} {\bibfnamefont {D.~S.}\ \bibnamefont {Sanchez}},
  \bibinfo {author} {\bibfnamefont {I.}~\bibnamefont {Belopolski}}, \bibinfo
  {author} {\bibfnamefont {N.}~\bibnamefont {Alidoust}}, \bibinfo {author}
  {\bibfnamefont {P.-J.}\ \bibnamefont {Chen}}, \bibinfo {author}
  {\bibfnamefont {G.}~\bibnamefont {Chang}}, \bibinfo {author} {\bibfnamefont
  {A.}~\bibnamefont {Bansil}}, \bibinfo {author} {\bibfnamefont {H.-T.}\
  \bibnamefont {Jeng}}, \bibinfo {author} {\bibfnamefont {H.}~\bibnamefont
  {Lin}},\ and\ \bibinfo {author} {\bibfnamefont {M.~Z.}\ \bibnamefont
  {Hasan}},\ }\bibfield  {title} {\bibinfo {title} {Drumhead surface states and
  topological nodal-line fermions in {T}l{T}a{S}e$_2$},\ }\href
  {https://doi.org/10.1103/PhysRevB.93.121113} {\bibfield  {journal} {\bibinfo
  {journal} {Phys. Rev. B}\ }\textbf {\bibinfo {volume} {93}},\ \bibinfo
  {pages} {121113} (\bibinfo {year} {2016}{\natexlab{b}})}\BibitemShut
  {NoStop}%
\bibitem [{\citenamefont {Bradlyn}\ \emph {et~al.}(2017)\citenamefont
  {Bradlyn}, \citenamefont {Elcoro}, \citenamefont {Cano}, \citenamefont
  {Vergniory}, \citenamefont {Wang}, \citenamefont {Felser}, \citenamefont
  {Aroyo},\ and\ \citenamefont {Bernevig}}]{bradlyn}%
  \BibitemOpen
  \bibfield  {author} {\bibinfo {author} {\bibfnamefont {B.}~\bibnamefont
  {Bradlyn}}, \bibinfo {author} {\bibfnamefont {L.}~\bibnamefont {Elcoro}},
  \bibinfo {author} {\bibfnamefont {J.}~\bibnamefont {Cano}}, \bibinfo {author}
  {\bibfnamefont {M.~G.}\ \bibnamefont {Vergniory}}, \bibinfo {author}
  {\bibfnamefont {Z.}~\bibnamefont {Wang}}, \bibinfo {author} {\bibfnamefont
  {C.}~\bibnamefont {Felser}}, \bibinfo {author} {\bibfnamefont {M.~I.}\
  \bibnamefont {Aroyo}},\ and\ \bibinfo {author} {\bibfnamefont {B.~A.}\
  \bibnamefont {Bernevig}},\ }\bibfield  {title} {\bibinfo {title} {Topological
  quantum chemistry},\ }\href {https://doi.org/10.1038/nature23268} {\bibfield
  {journal} {\bibinfo  {journal} {Nature}\ }\textbf {\bibinfo {volume} {547}},\
  \bibinfo {pages} {298} (\bibinfo {year} {2017})}\BibitemShut {NoStop}%
\bibitem [{\citenamefont {Cano}\ and\ \citenamefont {Bradlyn}(2021)}]{cano}%
  \BibitemOpen
  \bibfield  {author} {\bibinfo {author} {\bibfnamefont {J.}~\bibnamefont
  {Cano}}\ and\ \bibinfo {author} {\bibfnamefont {B.}~\bibnamefont {Bradlyn}},\
  }\bibfield  {title} {\bibinfo {title} {Band representations and topological
  quantum chemistry},\ }\href
  {https://doi.org/10.1146/annurev-conmatphys-041720-124134} {\bibfield
  {journal} {\bibinfo  {journal} {Annual Review of Condensed Matter Physics}\
  }\textbf {\bibinfo {volume} {12}},\ \bibinfo {pages} {225} (\bibinfo {year}
  {2021})}\BibitemShut {NoStop}%
\bibitem [{\citenamefont {Dzsaber}\ \emph {et~al.}(2022)\citenamefont
  {Dzsaber}, \citenamefont {Zocco}, \citenamefont {McCollam}, \citenamefont
  {Weickert}, \citenamefont {McDonald}, \citenamefont {Taupin}, \citenamefont
  {Eguchi}, \citenamefont {Yan}, \citenamefont {Prokofiev}, \citenamefont
  {Tang}, \citenamefont {Vlaar}, \citenamefont {Winter}, \citenamefont {Jaime},
  \citenamefont {Si},\ and\ \citenamefont {Paschen}}]{Dzsaber2}%
  \BibitemOpen
  \bibfield  {author} {\bibinfo {author} {\bibfnamefont {S.}~\bibnamefont
  {Dzsaber}}, \bibinfo {author} {\bibfnamefont {D.~A.}\ \bibnamefont {Zocco}},
  \bibinfo {author} {\bibfnamefont {A.}~\bibnamefont {McCollam}}, \bibinfo
  {author} {\bibfnamefont {F.}~\bibnamefont {Weickert}}, \bibinfo {author}
  {\bibfnamefont {R.}~\bibnamefont {McDonald}}, \bibinfo {author}
  {\bibfnamefont {M.}~\bibnamefont {Taupin}}, \bibinfo {author} {\bibfnamefont
  {G.}~\bibnamefont {Eguchi}}, \bibinfo {author} {\bibfnamefont
  {X.}~\bibnamefont {Yan}}, \bibinfo {author} {\bibfnamefont {A.}~\bibnamefont
  {Prokofiev}}, \bibinfo {author} {\bibfnamefont {L.~M.~K.}\ \bibnamefont
  {Tang}}, \bibinfo {author} {\bibfnamefont {B.}~\bibnamefont {Vlaar}},
  \bibinfo {author} {\bibfnamefont {L.~E.}\ \bibnamefont {Winter}}, \bibinfo
  {author} {\bibfnamefont {M.}~\bibnamefont {Jaime}}, \bibinfo {author}
  {\bibfnamefont {Q.}~\bibnamefont {Si}},\ and\ \bibinfo {author}
  {\bibfnamefont {S.}~\bibnamefont {Paschen}},\ }\bibfield  {title} {\bibinfo
  {title} {Control of electronic topology in a strongly correlated electron
  system},\ }\href {https://doi.org/10.1038/s41467-022-33369-8} {\bibfield
  {journal} {\bibinfo  {journal} {Nature Commun}\ }\textbf {\bibinfo {volume}
  {13}},\ \bibinfo {pages} {5729} (\bibinfo {year} {2022})}\BibitemShut
  {NoStop}%
\bibitem [{\citenamefont {Asaba}\ \emph {et~al.}(2020)\citenamefont {Asaba},
  \citenamefont {Su}, \citenamefont {Janoschek}, \citenamefont {Thompson},
  \citenamefont {Thomas}, \citenamefont {Bauer}, \citenamefont {Lin},\ and\
  \citenamefont {Ronning}}]{Asaba}%
  \BibitemOpen
  \bibfield  {author} {\bibinfo {author} {\bibfnamefont {T.}~\bibnamefont
  {Asaba}}, \bibinfo {author} {\bibfnamefont {Y.}~\bibnamefont {Su}}, \bibinfo
  {author} {\bibfnamefont {M.}~\bibnamefont {Janoschek}}, \bibinfo {author}
  {\bibfnamefont {J.~D.}\ \bibnamefont {Thompson}}, \bibinfo {author}
  {\bibfnamefont {S.~M.}\ \bibnamefont {Thomas}}, \bibinfo {author}
  {\bibfnamefont {E.~D.}\ \bibnamefont {Bauer}}, \bibinfo {author}
  {\bibfnamefont {S.-Z.}\ \bibnamefont {Lin}},\ and\ \bibinfo {author}
  {\bibfnamefont {F.}~\bibnamefont {Ronning}},\ }\bibfield  {title} {\bibinfo
  {title} {Large tunable anomalous hall effect in the kagome antiferromagnet
  {U}$_3${R}u$_4${A}l$_{12}$},\ }\href
  {https://doi.org/10.1103/PhysRevB.102.035127} {\bibfield  {journal} {\bibinfo
   {journal} {Phys. Rev. B}\ }\textbf {\bibinfo {volume} {102}},\ \bibinfo
  {pages} {035127} (\bibinfo {year} {2020})}\BibitemShut {NoStop}%
\bibitem [{\citenamefont {\ifmmode~\bar{O}\else \={O}\fi{}nuki}\ \emph
  {et~al.}(2023)\citenamefont {\ifmmode~\bar{O}\else \={O}\fi{}nuki},
  \citenamefont {Nakaima}, \citenamefont {Iha}, \citenamefont {Matsuda},
  \citenamefont {Hedo}, \citenamefont {Nakama}, \citenamefont {Aoki},
  \citenamefont {Nakamura}, \citenamefont {Nakashima}, \citenamefont {Amako},
  \citenamefont {Takeuchi},\ and\ \citenamefont {Matsuda}}]{Onuki}%
  \BibitemOpen
  \bibfield  {author} {\bibinfo {author} {\bibfnamefont {Y.}~\bibnamefont
  {\ifmmode~\bar{O}\else \={O}\fi{}nuki}}, \bibinfo {author} {\bibfnamefont
  {K.}~\bibnamefont {Nakaima}}, \bibinfo {author} {\bibfnamefont
  {W.}~\bibnamefont {Iha}}, \bibinfo {author} {\bibfnamefont {S.}~\bibnamefont
  {Matsuda}}, \bibinfo {author} {\bibfnamefont {M.}~\bibnamefont {Hedo}},
  \bibinfo {author} {\bibfnamefont {T.}~\bibnamefont {Nakama}}, \bibinfo
  {author} {\bibfnamefont {D.}~\bibnamefont {Aoki}}, \bibinfo {author}
  {\bibfnamefont {A.}~\bibnamefont {Nakamura}}, \bibinfo {author}
  {\bibfnamefont {M.}~\bibnamefont {Nakashima}}, \bibinfo {author}
  {\bibfnamefont {Y.}~\bibnamefont {Amako}}, \bibinfo {author} {\bibfnamefont
  {T.}~\bibnamefont {Takeuchi}},\ and\ \bibinfo {author} {\bibfnamefont
  {T.~D.}\ \bibnamefont {Matsuda}},\ }\bibfield  {title} {\bibinfo {title}
  {Anomalous hall effect in rare earth antiferromagnets with the hexagonal
  structures},\ }\href {https://doi.org/10.3938/NPSM.73.1054} {\bibfield
  {journal} {\bibinfo  {journal} {New Phys.: Sae Mulli}\ }\textbf {\bibinfo
  {volume} {73}},\ \bibinfo {pages} {1054} (\bibinfo {year}
  {2023})}\BibitemShut {NoStop}%
\bibitem [{\citenamefont {Kang}\ \emph {et~al.}(2020)\citenamefont {Kang},
  \citenamefont {Ye}, \citenamefont {Fang}, \citenamefont {You}, \citenamefont
  {Levitan}, \citenamefont {Han}, \citenamefont {Facio}, \citenamefont
  {Jozwiak}, \citenamefont {Bostwick}, \citenamefont {Rotenberg}, \citenamefont
  {Chan}, \citenamefont {McDonald}, \citenamefont {Graf}, \citenamefont
  {Kaznatcheev}, \citenamefont {Vescovo}, \citenamefont {Bell}, \citenamefont
  {Kaxiras}, \citenamefont {van~den Brink}, \citenamefont {Richter},
  \citenamefont {Prasad~Ghimire}, \citenamefont {Checkelsky},\ and\
  \citenamefont {Comin}}]{Kang}%
  \BibitemOpen
  \bibfield  {author} {\bibinfo {author} {\bibfnamefont {M.}~\bibnamefont
  {Kang}}, \bibinfo {author} {\bibfnamefont {L.}~\bibnamefont {Ye}}, \bibinfo
  {author} {\bibfnamefont {S.}~\bibnamefont {Fang}}, \bibinfo {author}
  {\bibfnamefont {J.-S.}\ \bibnamefont {You}}, \bibinfo {author} {\bibfnamefont
  {A.}~\bibnamefont {Levitan}}, \bibinfo {author} {\bibfnamefont
  {M.}~\bibnamefont {Han}}, \bibinfo {author} {\bibfnamefont {J.~I.}\
  \bibnamefont {Facio}}, \bibinfo {author} {\bibfnamefont {C.}~\bibnamefont
  {Jozwiak}}, \bibinfo {author} {\bibfnamefont {A.}~\bibnamefont {Bostwick}},
  \bibinfo {author} {\bibfnamefont {E.}~\bibnamefont {Rotenberg}}, \bibinfo
  {author} {\bibfnamefont {M.~K.}\ \bibnamefont {Chan}}, \bibinfo {author}
  {\bibfnamefont {R.~D.}\ \bibnamefont {McDonald}}, \bibinfo {author}
  {\bibfnamefont {D.}~\bibnamefont {Graf}}, \bibinfo {author} {\bibfnamefont
  {K.}~\bibnamefont {Kaznatcheev}}, \bibinfo {author} {\bibfnamefont
  {E.}~\bibnamefont {Vescovo}}, \bibinfo {author} {\bibfnamefont {D.~C.}\
  \bibnamefont {Bell}}, \bibinfo {author} {\bibfnamefont {E.}~\bibnamefont
  {Kaxiras}}, \bibinfo {author} {\bibfnamefont {J.}~\bibnamefont {van~den
  Brink}}, \bibinfo {author} {\bibfnamefont {M.}~\bibnamefont {Richter}},
  \bibinfo {author} {\bibfnamefont {M.}~\bibnamefont {Prasad~Ghimire}},
  \bibinfo {author} {\bibfnamefont {J.~G.}\ \bibnamefont {Checkelsky}},\ and\
  \bibinfo {author} {\bibfnamefont {R.}~\bibnamefont {Comin}},\ }\bibfield
  {title} {\bibinfo {title} {Dirac fermions and flat bands in the ideal kagome
  metal {F}e{S}n},\ }\href {https://doi.org/10.1038/s41563-019-0531-0}
  {\bibfield  {journal} {\bibinfo  {journal} {Nat. Mater.}\ }\textbf {\bibinfo
  {volume} {19}},\ \bibinfo {pages} {163–169} (\bibinfo {year}
  {2020})}\BibitemShut {NoStop}%
\bibitem [{\citenamefont {Chen}\ \emph {et~al.}(2022)\citenamefont {Chen},
  \citenamefont {Setty}, \citenamefont {Hu}, \citenamefont {Vergniory},
  \citenamefont {Grefe}, \citenamefont {Fischer}, \citenamefont {Yan},
  \citenamefont {Eguchi}, \citenamefont {Prokofiev}, \citenamefont {Paschen},
  \citenamefont {Cano},\ and\ \citenamefont {Si}}]{Chen2022}%
  \BibitemOpen
  \bibfield  {author} {\bibinfo {author} {\bibfnamefont {L.}~\bibnamefont
  {Chen}}, \bibinfo {author} {\bibfnamefont {C.}~\bibnamefont {Setty}},
  \bibinfo {author} {\bibfnamefont {H.}~\bibnamefont {Hu}}, \bibinfo {author}
  {\bibfnamefont {M.~G.}\ \bibnamefont {Vergniory}}, \bibinfo {author}
  {\bibfnamefont {S.~E.}\ \bibnamefont {Grefe}}, \bibinfo {author}
  {\bibfnamefont {L.}~\bibnamefont {Fischer}}, \bibinfo {author} {\bibfnamefont
  {X.}~\bibnamefont {Yan}}, \bibinfo {author} {\bibfnamefont {G.}~\bibnamefont
  {Eguchi}}, \bibinfo {author} {\bibfnamefont {A.}~\bibnamefont {Prokofiev}},
  \bibinfo {author} {\bibfnamefont {S.}~\bibnamefont {Paschen}}, \bibinfo
  {author} {\bibfnamefont {J.}~\bibnamefont {Cano}},\ and\ \bibinfo {author}
  {\bibfnamefont {Q.}~\bibnamefont {Si}},\ }\bibfield  {title} {\bibinfo
  {title} {Topological semimetal driven by strong correlations and crystalline
  symmetry},\ }\href
  {https://doi.org/https://doi.org/10.1038/s41567-022-01743-4} {\bibfield
  {journal} {\bibinfo  {journal} {Nature Phys}\ }\textbf {\bibinfo {volume}
  {18}},\ \bibinfo {pages} {1341} (\bibinfo {year} {2022})}\BibitemShut
  {NoStop}%
\bibitem [{\citenamefont {Simeth}\ \emph {et~al.}()\citenamefont {Simeth},
  \citenamefont {Hayami}, \citenamefont {Flury}, \citenamefont {Zaharko},
  \citenamefont {White}, \citenamefont {Su}, \citenamefont {Girod},
  \citenamefont {Francoual}, \citenamefont {Franz}, \citenamefont {Petricek},
  \citenamefont {Beauvois}, \citenamefont {Bartkowiak}, \citenamefont {Thomas},
  \citenamefont {Rosa}, \citenamefont {Lin},\ and\ \citenamefont
  {Janoschek}}]{Simeth2024}%
  \BibitemOpen
  \bibfield  {author} {\bibinfo {author} {\bibfnamefont {W.}~\bibnamefont
  {Simeth}}, \bibinfo {author} {\bibfnamefont {S.}~\bibnamefont {Hayami}},
  \bibinfo {author} {\bibfnamefont {S.}~\bibnamefont {Flury}}, \bibinfo
  {author} {\bibfnamefont {O.}~\bibnamefont {Zaharko}}, \bibinfo {author}
  {\bibfnamefont {J.~S.}\ \bibnamefont {White}}, \bibinfo {author}
  {\bibfnamefont {Y.}~\bibnamefont {Su}}, \bibinfo {author} {\bibfnamefont
  {C.}~\bibnamefont {Girod}}, \bibinfo {author} {\bibfnamefont
  {S.}~\bibnamefont {Francoual}}, \bibinfo {author} {\bibfnamefont
  {C.}~\bibnamefont {Franz}}, \bibinfo {author} {\bibfnamefont
  {V.}~\bibnamefont {Petricek}}, \bibinfo {author} {\bibfnamefont
  {K.}~\bibnamefont {Beauvois}}, \bibinfo {author} {\bibfnamefont
  {M.}~\bibnamefont {Bartkowiak}}, \bibinfo {author} {\bibfnamefont {S.~M.}\
  \bibnamefont {Thomas}}, \bibinfo {author} {\bibfnamefont {P.~F.~S.}\
  \bibnamefont {Rosa}}, \bibinfo {author} {\bibfnamefont {S.-Z.}\ \bibnamefont
  {Lin}},\ and\ \bibinfo {author} {\bibfnamefont {M.}~\bibnamefont
  {Janoschek}},\ }\bibfield  {title} {\bibinfo {title} {Skyrmion lattice order
  with alternating topological charges mediated by local
  {D}zyaloshinskii-{M}oriya interactions},\ }\href@noop {} {\bibinfo  {journal}
  {unpublished}\ }\BibitemShut {NoStop}%
\bibitem [{\citenamefont {Kaczorowski}\ \emph {et~al.}(1998)\citenamefont
  {Kaczorowski}, \citenamefont {Kruk}, \citenamefont {Sanchez}, \citenamefont
  {Malaman},\ and\ \citenamefont {Wastin}}]{Kaczorowski}%
  \BibitemOpen
\bibfield  {journal} {  }\bibfield  {author} {\bibinfo {author} {\bibfnamefont
  {D.}~\bibnamefont {Kaczorowski}}, \bibinfo {author} {\bibfnamefont
  {R.}~\bibnamefont {Kruk}}, \bibinfo {author} {\bibfnamefont {J.~P.}\
  \bibnamefont {Sanchez}}, \bibinfo {author} {\bibfnamefont {B.}~\bibnamefont
  {Malaman}},\ and\ \bibinfo {author} {\bibfnamefont {F.}~\bibnamefont
  {Wastin}},\ }\bibfield  {title} {\bibinfo {title} {Magnetic and electronic
  properties of ternary uranium antimonides {U}{T}{S}b$_{2}$ ({T}=3d -, 4d-,
  5d- electron transition metal)},\ }\href
  {https://doi.org/10.1103/PhysRevB.58.9227} {\bibfield  {journal} {\bibinfo
  {journal} {Phys. Rev. B}\ }\textbf {\bibinfo {volume} {58}},\ \bibinfo
  {pages} {9227} (\bibinfo {year} {1998})}\BibitemShut {NoStop}%
\bibitem [{\citenamefont {Kaczorowski}(1992)}]{KACZOROWSKI1992333}%
  \BibitemOpen
  \bibfield  {author} {\bibinfo {author} {\bibfnamefont {D.}~\bibnamefont
  {Kaczorowski}},\ }\bibfield  {title} {\bibinfo {title} {Structural and
  magnetic properties of some new uranium ternary pnictides: {U}{T}{X}$_2$ ({T}
  { F}e, {C}o, {N}i, {C}u; {X} {P}, {A}s, {S}b, {B}i)},\ }\href
  {https://doi.org/https://doi.org/10.1016/0925-8388(92)90020-A} {\bibfield
  {journal} {\bibinfo  {journal} {Journal of Alloys and Compounds}\ }\textbf
  {\bibinfo {volume} {186}},\ \bibinfo {pages} {333} (\bibinfo {year}
  {1992})}\BibitemShut {NoStop}%
\bibitem [{\citenamefont {Ikeda}\ \emph {et~al.}(2004)\citenamefont {Ikeda},
  \citenamefont {Matsuda}, \citenamefont {Galatanu}, \citenamefont {Yamamoto},
  \citenamefont {Haga},\ and\ \citenamefont {Ōnuki}}]{IKEDA200462}%
  \BibitemOpen
  \bibfield  {author} {\bibinfo {author} {\bibfnamefont {S.}~\bibnamefont
  {Ikeda}}, \bibinfo {author} {\bibfnamefont {T.}~\bibnamefont {Matsuda}},
  \bibinfo {author} {\bibfnamefont {A.}~\bibnamefont {Galatanu}}, \bibinfo
  {author} {\bibfnamefont {E.}~\bibnamefont {Yamamoto}}, \bibinfo {author}
  {\bibfnamefont {Y.}~\bibnamefont {Haga}},\ and\ \bibinfo {author}
  {\bibfnamefont {Y.}~\bibnamefont {Ōnuki}},\ }\bibfield  {title} {\bibinfo
  {title} {Single crystal growth and magnetic property of {U}{N}i{S}b$_2$},\
  }\href {https://doi.org/https://doi.org/10.1016/j.jmmm.2003.11.055}
  {\bibfield  {journal} {\bibinfo  {journal} {Journal of Magnetism and Magnetic
  Materials}\ }\textbf {\bibinfo {volume} {272-276}},\ \bibinfo {pages} {62}
  (\bibinfo {year} {2004})},\ \bibinfo {note} {proceedings of the International
  Conference on Magnetism (ICM 2003)}\BibitemShut {NoStop}%
\bibitem [{\citenamefont {Rosa}\ \emph
  {et~al.}(2015{\natexlab{a}})\citenamefont {Rosa}, \citenamefont {Luo},
  \citenamefont {Bauer}, \citenamefont {Thompson}, \citenamefont {Pagliuso},\
  and\ \citenamefont {Fisk}}]{PhysRevB.92.104425}%
  \BibitemOpen
  \bibfield  {author} {\bibinfo {author} {\bibfnamefont {P.~F.~S.}\
  \bibnamefont {Rosa}}, \bibinfo {author} {\bibfnamefont {Y.}~\bibnamefont
  {Luo}}, \bibinfo {author} {\bibfnamefont {E.~D.}\ \bibnamefont {Bauer}},
  \bibinfo {author} {\bibfnamefont {J.~D.}\ \bibnamefont {Thompson}}, \bibinfo
  {author} {\bibfnamefont {P.~G.}\ \bibnamefont {Pagliuso}},\ and\ \bibinfo
  {author} {\bibfnamefont {Z.}~\bibnamefont {Fisk}},\ }\bibfield  {title}
  {\bibinfo {title} {Ferromagnetic kondo behavior in {U}{A}u{B}i$_{2}$ single
  crystals},\ }\href {https://doi.org/10.1103/PhysRevB.92.104425} {\bibfield
  {journal} {\bibinfo  {journal} {Phys. Rev. B}\ }\textbf {\bibinfo {volume}
  {92}},\ \bibinfo {pages} {104425} (\bibinfo {year}
  {2015}{\natexlab{a}})}\BibitemShut {NoStop}%
\bibitem [{\citenamefont {Freitas}\ \emph {et~al.}()\citenamefont {Freitas},
  \citenamefont {Girod}, \citenamefont {Yahne}, \citenamefont {Simeth},
  \citenamefont {Kengle}, \citenamefont {Carneiro}, \citenamefont {Pagliuso},
  \citenamefont {Thompson}, \citenamefont {Janoschek}, \citenamefont {Thomas},\
  and\ \citenamefont {Rosa}}]{Gabriel2024}%
  \BibitemOpen
  \bibfield  {author} {\bibinfo {author} {\bibfnamefont {G.~S.}\ \bibnamefont
  {Freitas}}, \bibinfo {author} {\bibfnamefont {C.}~\bibnamefont {Girod}},
  \bibinfo {author} {\bibfnamefont {D.~R.}\ \bibnamefont {Yahne}}, \bibinfo
  {author} {\bibfnamefont {W.}~\bibnamefont {Simeth}}, \bibinfo {author}
  {\bibfnamefont {C.~S.}\ \bibnamefont {Kengle}}, \bibinfo {author}
  {\bibfnamefont {F.~B.}\ \bibnamefont {Carneiro}}, \bibinfo {author}
  {\bibfnamefont {P.~G.}\ \bibnamefont {Pagliuso}}, \bibinfo {author}
  {\bibfnamefont {J.~D.}\ \bibnamefont {Thompson}}, \bibinfo {author}
  {\bibfnamefont {M.}~\bibnamefont {Janoschek}}, \bibinfo {author}
  {\bibfnamefont {S.~M.}\ \bibnamefont {Thomas}},\ and\ \bibinfo {author}
  {\bibfnamefont {P.~F.~S.}\ \bibnamefont {Rosa}},\ }\bibfield  {title}
  {\bibinfo {title} {Magnetic devil's staircase in {U}{A}g{B}i$_{2}$},\
  }\href@noop {} {\bibinfo  {journal} {unpublished}\ }\BibitemShut {NoStop}%
\bibitem [{\citenamefont {Rosa}\ and\ \citenamefont {Fisk}(2019)}]{Rosa2019}%
  \BibitemOpen
\bibfield  {journal} {  }\bibfield  {author} {\bibinfo {author} {\bibfnamefont
  {P.~F.~S.}\ \bibnamefont {Rosa}}\ and\ \bibinfo {author} {\bibfnamefont
  {Z.}~\bibnamefont {Fisk}},\ }\bibinfo {title} {Flux methods for growth of
  intermetallic single crystals},\ in\ \href
  {https://doi.org/doi:10.1515/9783110496789-003} {\emph {\bibinfo {booktitle}
  {Crystal Growth of Intermetallics}}},\ \bibinfo {editor} {edited by\ \bibinfo
  {editor} {\bibfnamefont {P.}~\bibnamefont {Gille}}\ and\ \bibinfo {editor}
  {\bibfnamefont {Y.}~\bibnamefont {Grin}}}\ (\bibinfo  {publisher} {De
  Gruyter},\ \bibinfo {address} {Berlin, Boston},\ \bibinfo {year} {2019})\
  pp.\ \bibinfo {pages} {49--60}\BibitemShut {NoStop}%
\bibitem [{\citenamefont {Sheldrick}(2015)}]{Sheldrick2015acta}%
  \BibitemOpen
  \bibfield  {author} {\bibinfo {author} {\bibfnamefont {G.~M.}\ \bibnamefont
  {Sheldrick}},\ }\bibfield  {title} {\bibinfo {title} {Shelxt--integrated
  space-group and crystal-structure determination},\ }\href
  {https://doi.org/10.1107/S2053273314026370} {\bibfield  {journal} {\bibinfo
  {journal} {Acta Crystallographica Section A: Foundations and Advances}\
  }\textbf {\bibinfo {volume} {71}},\ \bibinfo {pages} {3} (\bibinfo {year}
  {2015})}\BibitemShut {NoStop}%
\bibitem [{\citenamefont {Tkachuk}\ and\ \citenamefont
  {Mar}(2004)}]{tkachuk2004cerium}%
  \BibitemOpen
  \bibfield  {author} {\bibinfo {author} {\bibfnamefont {A.~V.}\ \bibnamefont
  {Tkachuk}}\ and\ \bibinfo {author} {\bibfnamefont {A.}~\bibnamefont {Mar}},\
  }\bibfield  {title} {\bibinfo {title} {{Cerium cadmium diantimonide,
  CeCd$_{0.660}$Sb$_2$}},\ }\href {https://doi.org/10.1107/S1600536804012814}
  {\bibfield  {journal} {\bibinfo  {journal} {Acta Crystallographica Section E:
  Structure Reports Online}\ }\textbf {\bibinfo {volume} {60}},\ \bibinfo
  {pages} {i82} (\bibinfo {year} {2004})}\BibitemShut {NoStop}%
\bibitem [{\citenamefont {Sharma}\ and\ \citenamefont
  {Thamizhavel}(2023)}]{Sharma2023PRB}%
  \BibitemOpen
  \bibfield  {author} {\bibinfo {author} {\bibfnamefont {V.}~\bibnamefont
  {Sharma}}\ and\ \bibinfo {author} {\bibfnamefont {A.}~\bibnamefont
  {Thamizhavel}},\ }\bibfield  {title} {\bibinfo {title} {{Anisotropic magnetic
  properties of RCd$_{1-\delta}$Sb$_{2}$ (R = Ce$-$Nd) single crystals}},\
  }\href {https://doi.org/10.1103/PhysRevB.108.214403} {\bibfield  {journal}
  {\bibinfo  {journal} {Phys. Rev. B}\ }\textbf {\bibinfo {volume} {108}},\
  \bibinfo {pages} {214403} (\bibinfo {year} {2023})}\BibitemShut {NoStop}%
\bibitem [{\citenamefont {Rosa}\ \emph
  {et~al.}(2015{\natexlab{b}})\citenamefont {Rosa}, \citenamefont {Bourg},
  \citenamefont {Jesus}, \citenamefont {Pagliuso},\ and\ \citenamefont
  {Fisk}}]{Rosa2015PRB}%
  \BibitemOpen
  \bibfield  {author} {\bibinfo {author} {\bibfnamefont {P.~F.~S.}\
  \bibnamefont {Rosa}}, \bibinfo {author} {\bibfnamefont {R.~J.}\ \bibnamefont
  {Bourg}}, \bibinfo {author} {\bibfnamefont {C.~B.~R.}\ \bibnamefont {Jesus}},
  \bibinfo {author} {\bibfnamefont {P.~G.}\ \bibnamefont {Pagliuso}},\ and\
  \bibinfo {author} {\bibfnamefont {Z.}~\bibnamefont {Fisk}},\ }\bibfield
  {title} {\bibinfo {title} {{Role of dimensionality in the Kondo CeTX$_{2}$
  family: The case of CeCd$_{0.7}$Sb$_{2}$}},\ }\href
  {https://doi.org/10.1103/PhysRevB.92.134421} {\bibfield  {journal} {\bibinfo
  {journal} {Phys. Rev. B}\ }\textbf {\bibinfo {volume} {92}},\ \bibinfo
  {pages} {134421} (\bibinfo {year} {2015}{\natexlab{b}})}\BibitemShut
  {NoStop}%
\bibitem [{\citenamefont {Piva}\ \emph {et~al.}(2021)\citenamefont {Piva},
  \citenamefont {Rahn}, \citenamefont {Thomas}, \citenamefont {Scott},
  \citenamefont {Pagliuso}, \citenamefont {Thompson}, \citenamefont {Schoop},
  \citenamefont {Ronning},\ and\ \citenamefont {Rosa}}]{Piva2021ACS}%
  \BibitemOpen
  \bibfield  {author} {\bibinfo {author} {\bibfnamefont {M.~M.}\ \bibnamefont
  {Piva}}, \bibinfo {author} {\bibfnamefont {M.~C.}\ \bibnamefont {Rahn}},
  \bibinfo {author} {\bibfnamefont {S.~M.}\ \bibnamefont {Thomas}}, \bibinfo
  {author} {\bibfnamefont {B.~L.}\ \bibnamefont {Scott}}, \bibinfo {author}
  {\bibfnamefont {P.~G.}\ \bibnamefont {Pagliuso}}, \bibinfo {author}
  {\bibfnamefont {J.~D.}\ \bibnamefont {Thompson}}, \bibinfo {author}
  {\bibfnamefont {L.~M.}\ \bibnamefont {Schoop}}, \bibinfo {author}
  {\bibfnamefont {F.}~\bibnamefont {Ronning}},\ and\ \bibinfo {author}
  {\bibfnamefont {P.~F.}\ \bibnamefont {Rosa}},\ }\bibfield  {title} {\bibinfo
  {title} {{Robust narrow-gap semiconducting behavior in square-net
  La$_3$Cd$_2$As$_6$}},\ }\href {https://doi.org/10.1021/acs.chemmater.1c00797}
  {\bibfield  {journal} {\bibinfo  {journal} {Chemistry of Materials}\ }\textbf
  {\bibinfo {volume} {33}},\ \bibinfo {pages} {4122} (\bibinfo {year}
  {2021})}\BibitemShut {NoStop}%
\bibitem [{\citenamefont {Lei}\ \emph {et~al.}(2019)\citenamefont {Lei},
  \citenamefont {Duppel}, \citenamefont {Lippmann}, \citenamefont {Nuss},
  \citenamefont {Lotsch},\ and\ \citenamefont {Schoop}}]{Lei}%
  \BibitemOpen
  \bibfield  {author} {\bibinfo {author} {\bibfnamefont {S.}~\bibnamefont
  {Lei}}, \bibinfo {author} {\bibfnamefont {V.}~\bibnamefont {Duppel}},
  \bibinfo {author} {\bibfnamefont {J.~M.}\ \bibnamefont {Lippmann}}, \bibinfo
  {author} {\bibfnamefont {J.}~\bibnamefont {Nuss}}, \bibinfo {author}
  {\bibfnamefont {B.~V.}\ \bibnamefont {Lotsch}},\ and\ \bibinfo {author}
  {\bibfnamefont {L.~M.}\ \bibnamefont {Schoop}},\ }\bibfield  {title}
  {\bibinfo {title} {Charge density waves and magnetism in topological
  semimetal candidates {G}d{S}b$_x${T}e$_{2-x-\delta}$},\ }\href
  {https://doi.org/https://doi.org/10.1002/qute.201900045} {\bibfield
  {journal} {\bibinfo  {journal} {Advanced Quantum Technologies}\ }\textbf
  {\bibinfo {volume} {2}},\ \bibinfo {pages} {1900045} (\bibinfo {year}
  {2019})}\BibitemShut {NoStop}%
\bibitem [{\citenamefont {Patschke}\ and\ \citenamefont
  {Kanatzidis}(2002)}]{Patschke}%
  \BibitemOpen
  \bibfield  {author} {\bibinfo {author} {\bibfnamefont {R.}~\bibnamefont
  {Patschke}}\ and\ \bibinfo {author} {\bibfnamefont {M.~G.}\ \bibnamefont
  {Kanatzidis}},\ }\bibfield  {title} {\bibinfo {title} {Polytelluride
  compounds containing distorted nets of tellurium},\ }\href
  {https://doi.org/10.1039/B201162J} {\bibfield  {journal} {\bibinfo  {journal}
  {Phys. Chem. Chem. Phys.}\ }\textbf {\bibinfo {volume} {4}},\ \bibinfo
  {pages} {3266} (\bibinfo {year} {2002})}\BibitemShut {NoStop}%
\bibitem [{\citenamefont {DiMasi}\ \emph {et~al.}(1996)\citenamefont {DiMasi},
  \citenamefont {Foran}, \citenamefont {Aronson},\ and\ \citenamefont
  {Lee}}]{DiMasi}%
  \BibitemOpen
  \bibfield  {author} {\bibinfo {author} {\bibfnamefont {E.}~\bibnamefont
  {DiMasi}}, \bibinfo {author} {\bibfnamefont {B.}~\bibnamefont {Foran}},
  \bibinfo {author} {\bibfnamefont {M.~C.}\ \bibnamefont {Aronson}},\ and\
  \bibinfo {author} {\bibfnamefont {S.}~\bibnamefont {Lee}},\ }\bibfield
  {title} {\bibinfo {title} {Stability of charge-density waves under continuous
  variation of band filling in {L}a{T}e$_{2-x}${S}b$_{x}$ (0$<$x$<$1)},\ }\href
  {https://doi.org/10.1103/PhysRevB.54.13587} {\bibfield  {journal} {\bibinfo
  {journal} {Phys. Rev. B}\ }\textbf {\bibinfo {volume} {54}},\ \bibinfo
  {pages} {13587} (\bibinfo {year} {1996})}\BibitemShut {NoStop}%
\bibitem [{\citenamefont {Troć}(1987)}]{TROC198767}%
  \BibitemOpen
  \bibfield  {author} {\bibinfo {author} {\bibfnamefont {R.}~\bibnamefont
  {Troć}},\ }\bibfield  {title} {\bibinfo {title} {Structural and magnetic
  properties of the tetragonal actinide compounds},\ }\href
  {https://doi.org/https://doi.org/10.1016/S0020-1693(00)81052-X} {\bibfield
  {journal} {\bibinfo  {journal} {Inorganica Chimica Acta}\ }\textbf {\bibinfo
  {volume} {140}},\ \bibinfo {pages} {67} (\bibinfo {year} {1987})}\BibitemShut
  {NoStop}%
\bibitem [{\citenamefont {Bukowski}\ \emph {et~al.}(2004)\citenamefont
  {Bukowski}, \citenamefont {Tran}, \citenamefont {Stepień-Damm},\ and\
  \citenamefont {Troć}}]{BUKOWSKI20043934}%
  \BibitemOpen
  \bibfield  {author} {\bibinfo {author} {\bibfnamefont {Z.}~\bibnamefont
  {Bukowski}}, \bibinfo {author} {\bibfnamefont {V.}~\bibnamefont {Tran}},
  \bibinfo {author} {\bibfnamefont {J.}~\bibnamefont {Stepień-Damm}},\ and\
  \bibinfo {author} {\bibfnamefont {R.}~\bibnamefont {Troć}},\ }\bibfield
  {title} {\bibinfo {title} {Single crystal growth, crystal structure
  characterization and magnetic properties of {U}{C}o$_{0.5}${S}b$_2$},\ }\href
  {https://doi.org/https://doi.org/10.1016/j.jssc.2004.08.002} {\bibfield
  {journal} {\bibinfo  {journal} {Journal of Solid State Chemistry}\ }\textbf
  {\bibinfo {volume} {177}},\ \bibinfo {pages} {3934} (\bibinfo {year}
  {2004})}\BibitemShut {NoStop}%
\bibitem [{\citenamefont {Hayami}\ \emph {et~al.}(2017)\citenamefont {Hayami},
  \citenamefont {Ozawa},\ and\ \citenamefont {Motome}}]{Hayami}%
  \BibitemOpen
  \bibfield  {author} {\bibinfo {author} {\bibfnamefont {S.}~\bibnamefont
  {Hayami}}, \bibinfo {author} {\bibfnamefont {R.}~\bibnamefont {Ozawa}},\ and\
  \bibinfo {author} {\bibfnamefont {Y.}~\bibnamefont {Motome}},\ }\bibfield
  {title} {\bibinfo {title} {Effective bilinear-biquadratic model for
  noncoplanar ordering in itinerant magnets},\ }\href
  {https://doi.org/10.1103/PhysRevB.95.224424} {\bibfield  {journal} {\bibinfo
  {journal} {Phys. Rev. B}\ }\textbf {\bibinfo {volume} {95}},\ \bibinfo
  {pages} {224424} (\bibinfo {year} {2017})}\BibitemShut {NoStop}%
\bibitem [{\citenamefont {Seo}\ \emph {et~al.}(2021)\citenamefont {Seo},
  \citenamefont {Hayami}, \citenamefont {Su}, \citenamefont {Thomas},
  \citenamefont {Ronning}, \citenamefont {Bauer}, \citenamefont {Thompson},\
  and\ \citenamefont {Rosa}}]{Seo}%
  \BibitemOpen
  \bibfield  {author} {\bibinfo {author} {\bibfnamefont {S.}~\bibnamefont
  {Seo}}, \bibinfo {author} {\bibfnamefont {S.}~\bibnamefont {Hayami}},
  \bibinfo {author} {\bibfnamefont {Y.}~\bibnamefont {Su}}, \bibinfo {author}
  {\bibfnamefont {S.~M.}\ \bibnamefont {Thomas}}, \bibinfo {author}
  {\bibfnamefont {F.}~\bibnamefont {Ronning}}, \bibinfo {author} {\bibfnamefont
  {E.~D.}\ \bibnamefont {Bauer}}, \bibinfo {author} {\bibfnamefont {S.-Z.}\
  \bibnamefont {Thompson}, \bibfnamefont {Joe D.and~Lin}},\ and\ \bibinfo
  {author} {\bibfnamefont {P.~F.~S.}\ \bibnamefont {Rosa}},\ }\bibfield
  {title} {\bibinfo {title} {Spin-texture-driven electrical transport in
  multi-{Q} antiferromagnets},\ }\href
  {https://doi.org/10.1038/s42005-021-00558-8} {\bibfield  {journal} {\bibinfo
  {journal} {Commun Phys}\ }\textbf {\bibinfo {volume} {4}},\ \bibinfo {pages}
  {58} (\bibinfo {year} {2021})}\BibitemShut {NoStop}%
\bibitem [{\citenamefont {Fisher}\ and\ \citenamefont {Langer}(1968)}]{Fisher}%
  \BibitemOpen
  \bibfield  {author} {\bibinfo {author} {\bibfnamefont {M.~E.}\ \bibnamefont
  {Fisher}}\ and\ \bibinfo {author} {\bibfnamefont {J.~S.}\ \bibnamefont
  {Langer}},\ }\bibfield  {title} {\bibinfo {title} {Resistive anomalies at
  magnetic critical points},\ }\href
  {https://doi.org/10.1103/PhysRevLett.20.665} {\bibfield  {journal} {\bibinfo
  {journal} {Phys. Rev. Lett.}\ }\textbf {\bibinfo {volume} {20}},\ \bibinfo
  {pages} {665} (\bibinfo {year} {1968})}\BibitemShut {NoStop}%
\bibitem [{\citenamefont {Zumsteg}\ and\ \citenamefont
  {Parks}(1970)}]{Zumsteg}%
  \BibitemOpen
  \bibfield  {author} {\bibinfo {author} {\bibfnamefont {F.~C.}\ \bibnamefont
  {Zumsteg}}\ and\ \bibinfo {author} {\bibfnamefont {R.~D.}\ \bibnamefont
  {Parks}},\ }\bibfield  {title} {\bibinfo {title} {Electrical {R}esistivity of
  {N}ickel {N}ear the {C}urie {P}oint},\ }\href
  {https://doi.org/10.1103/PhysRevLett.24.520} {\bibfield  {journal} {\bibinfo
  {journal} {Phys. Rev. Lett.}\ }\textbf {\bibinfo {volume} {24}},\ \bibinfo
  {pages} {520} (\bibinfo {year} {1970})}\BibitemShut {NoStop}%
\bibitem [{\citenamefont {Jeong}\ \emph {et~al.}(1998)\citenamefont {Jeong},
  \citenamefont {Park}, \citenamefont {Koo},\ and\ \citenamefont {Lee}}]{Yoon}%
  \BibitemOpen
  \bibfield  {author} {\bibinfo {author} {\bibfnamefont {Y.-H.}\ \bibnamefont
  {Jeong}}, \bibinfo {author} {\bibfnamefont {S.}~\bibnamefont {Park}},
  \bibinfo {author} {\bibfnamefont {T.}~\bibnamefont {Koo}},\ and\ \bibinfo
  {author} {\bibfnamefont {K.-B.}\ \bibnamefont {Lee}},\ }\bibfield  {title}
  {\bibinfo {title} {Fisher–langer relation and scaling in the specific heat
  and resistivity of {L}a$_{0.7}${C}a$_{0.3}${M}n{O}$_3$},\ }\href
  {https://doi.org/https://doi.org/10.1016/S0167-2738(98)00046-0} {\bibfield
  {journal} {\bibinfo  {journal} {Solid State Ionics}\ }\textbf {\bibinfo
  {volume} {108}},\ \bibinfo {pages} {249} (\bibinfo {year}
  {1998})}\BibitemShut {NoStop}%
\bibitem [{\citenamefont {Mackintosh}(1962)}]{Mackintosh}%
  \BibitemOpen
  \bibfield  {author} {\bibinfo {author} {\bibfnamefont {A.~R.}\ \bibnamefont
  {Mackintosh}},\ }\bibfield  {title} {\bibinfo {title} {Magnetic ordering and
  the electronic structure of rare-earth metals},\ }\href
  {https://doi.org/10.1103/PhysRevLett.9.90} {\bibfield  {journal} {\bibinfo
  {journal} {Phys. Rev. Lett.}\ }\textbf {\bibinfo {volume} {9}},\ \bibinfo
  {pages} {90} (\bibinfo {year} {1962})}\BibitemShut {NoStop}%
\bibitem [{\citenamefont {Elliott}\ and\ \citenamefont
  {Wedgwood}(1963)}]{Elliott}%
  \BibitemOpen
  \bibfield  {author} {\bibinfo {author} {\bibfnamefont {R.~J.}\ \bibnamefont
  {Elliott}}\ and\ \bibinfo {author} {\bibfnamefont {F.~A.}\ \bibnamefont
  {Wedgwood}},\ }\bibfield  {title} {\bibinfo {title} {Theory of the resistance
  of the rare earth metals},\ }\href
  {https://doi.org/10.1088/0370-1328/81/5/308} {\bibfield  {journal} {\bibinfo
  {journal} {Proceedings of the Physical Society}\ }\textbf {\bibinfo {volume}
  {81}},\ \bibinfo {pages} {846} (\bibinfo {year} {1963})}\BibitemShut
  {NoStop}%
\bibitem [{\citenamefont {Becker}\ \emph {et~al.}(1997)\citenamefont {Becker},
  \citenamefont {Ramakrishnan}, \citenamefont {Menovsky}, \citenamefont
  {Nieuwenhuys},\ and\ \citenamefont {Mydosh}}]{Becker}%
  \BibitemOpen
  \bibfield  {author} {\bibinfo {author} {\bibfnamefont {B.}~\bibnamefont
  {Becker}}, \bibinfo {author} {\bibfnamefont {S.}~\bibnamefont
  {Ramakrishnan}}, \bibinfo {author} {\bibfnamefont {A.~A.}\ \bibnamefont
  {Menovsky}}, \bibinfo {author} {\bibfnamefont {G.~J.}\ \bibnamefont
  {Nieuwenhuys}},\ and\ \bibinfo {author} {\bibfnamefont {J.~A.}\ \bibnamefont
  {Mydosh}},\ }\bibfield  {title} {\bibinfo {title} {Unusual ordering behavior
  in single-crystal {U}$_{2}${R}h$_{3}${S}i$_{5}$},\ }\href
  {https://doi.org/10.1103/PhysRevLett.78.1347} {\bibfield  {journal} {\bibinfo
   {journal} {Phys. Rev. Lett.}\ }\textbf {\bibinfo {volume} {78}},\ \bibinfo
  {pages} {1347} (\bibinfo {year} {1997})}\BibitemShut {NoStop}%
\bibitem [{\citenamefont {Onimaru}\ \emph {et~al.}(2008)\citenamefont
  {Onimaru}, \citenamefont {F.~Inoue}, \citenamefont {Shigetoh}, \citenamefont
  {Umeo}, \citenamefont {Kubo}, \citenamefont {A.~Ribeiro}, \citenamefont
  {Ishida}, \citenamefont {A.~Avila}, \citenamefont {Ohoyama}, \citenamefont
  {Sera},\ and\ \citenamefont {Takabatake}}]{Onimaru}%
  \BibitemOpen
  \bibfield  {author} {\bibinfo {author} {\bibfnamefont {T.}~\bibnamefont
  {Onimaru}}, \bibinfo {author} {\bibfnamefont {Y.}~\bibnamefont {F.~Inoue}},
  \bibinfo {author} {\bibfnamefont {K.}~\bibnamefont {Shigetoh}}, \bibinfo
  {author} {\bibfnamefont {K.}~\bibnamefont {Umeo}}, \bibinfo {author}
  {\bibfnamefont {H.}~\bibnamefont {Kubo}}, \bibinfo {author} {\bibfnamefont
  {R.}~\bibnamefont {A.~Ribeiro}}, \bibinfo {author} {\bibfnamefont
  {A.}~\bibnamefont {Ishida}}, \bibinfo {author} {\bibfnamefont
  {M.}~\bibnamefont {A.~Avila}}, \bibinfo {author} {\bibfnamefont
  {K.}~\bibnamefont {Ohoyama}}, \bibinfo {author} {\bibfnamefont
  {M.}~\bibnamefont {Sera}},\ and\ \bibinfo {author} {\bibfnamefont
  {T.}~\bibnamefont {Takabatake}},\ }\bibfield  {title} {\bibinfo {title}
  {Giant uniaxial anisotropy in the magnetic and transport properties of
  {C}e{P}d$_5${A}l$_2$},\ }\href {https://doi.org/10.1143/JPSJ.77.074708}
  {\bibfield  {journal} {\bibinfo  {journal} {Journal of the Physical Society
  of Japan}\ }\textbf {\bibinfo {volume} {77}},\ \bibinfo {pages} {074708}
  (\bibinfo {year} {2008})}\BibitemShut {NoStop}%
\bibitem [{\citenamefont {Feng}\ \emph {et~al.}(2013)\citenamefont {Feng},
  \citenamefont {Wang}, \citenamefont {Silevitch}, \citenamefont {Mihaila},
  \citenamefont {Kim}, \citenamefont {Yan}, \citenamefont {Schulze},
  \citenamefont {Woo}, \citenamefont {Palmer}, \citenamefont {Ren},
  \citenamefont {van Wezel}, \citenamefont {Littlewood},\ and\ \citenamefont
  {Rosenbaum}}]{Yejun}%
  \BibitemOpen
  \bibfield  {author} {\bibinfo {author} {\bibfnamefont {Y.}~\bibnamefont
  {Feng}}, \bibinfo {author} {\bibfnamefont {J.}~\bibnamefont {Wang}}, \bibinfo
  {author} {\bibfnamefont {D.~M.}\ \bibnamefont {Silevitch}}, \bibinfo {author}
  {\bibfnamefont {B.}~\bibnamefont {Mihaila}}, \bibinfo {author} {\bibfnamefont
  {J.~W.}\ \bibnamefont {Kim}}, \bibinfo {author} {\bibfnamefont {J.-Q.}\
  \bibnamefont {Yan}}, \bibinfo {author} {\bibfnamefont {R.~K.}\ \bibnamefont
  {Schulze}}, \bibinfo {author} {\bibfnamefont {N.}~\bibnamefont {Woo}},
  \bibinfo {author} {\bibfnamefont {A.}~\bibnamefont {Palmer}}, \bibinfo
  {author} {\bibfnamefont {Y.}~\bibnamefont {Ren}}, \bibinfo {author}
  {\bibfnamefont {J.}~\bibnamefont {van Wezel}}, \bibinfo {author}
  {\bibfnamefont {P.~B.}\ \bibnamefont {Littlewood}},\ and\ \bibinfo {author}
  {\bibfnamefont {T.~F.}\ \bibnamefont {Rosenbaum}},\ }\bibfield  {title}
  {\bibinfo {title} {Incommensurate antiferromagnetism in a pure spin system
  via cooperative organization of local and itinerant moments},\ }\href
  {https://doi.org/10.1073/pnas.1217292110} {\bibfield  {journal} {\bibinfo
  {journal} {Proceedings of the National Academy of Sciences}\ }\textbf
  {\bibinfo {volume} {110}},\ \bibinfo {pages} {3287} (\bibinfo {year}
  {2013})}\BibitemShut {NoStop}%
\bibitem [{\citenamefont {Ru}\ \emph {et~al.}(2008)\citenamefont {Ru},
  \citenamefont {Chu},\ and\ \citenamefont {Fisher}}]{Ru}%
  \BibitemOpen
  \bibfield  {author} {\bibinfo {author} {\bibfnamefont {N.}~\bibnamefont
  {Ru}}, \bibinfo {author} {\bibfnamefont {J.-H.}\ \bibnamefont {Chu}},\ and\
  \bibinfo {author} {\bibfnamefont {I.~R.}\ \bibnamefont {Fisher}},\ }\bibfield
   {title} {\bibinfo {title} {Magnetic properties of the charge density wave
  compounds {R}{T}e$_{3}$ ({R}={Y}, {L}a, {C}e, {P}r, {N}d, {S}m, {G}d, {T}b,
  {D}y, {H}o, {E}r, and {T}m)},\ }\href
  {https://doi.org/10.1103/PhysRevB.78.012410} {\bibfield  {journal} {\bibinfo
  {journal} {Phys. Rev. B}\ }\textbf {\bibinfo {volume} {78}},\ \bibinfo
  {pages} {012410} (\bibinfo {year} {2008})}\BibitemShut {NoStop}%
\bibitem [{\citenamefont {Ajeesh}\ \emph {et~al.}(2022)\citenamefont {Ajeesh},
  \citenamefont {Thomas}, \citenamefont {Kushwaha}, \citenamefont {Bauer},
  \citenamefont {Ronning}, \citenamefont {Thompson}, \citenamefont {Harrison},\
  and\ \citenamefont {Rosa}}]{Ajeesh}%
  \BibitemOpen
  \bibfield  {author} {\bibinfo {author} {\bibfnamefont {M.~O.}\ \bibnamefont
  {Ajeesh}}, \bibinfo {author} {\bibfnamefont {S.~M.}\ \bibnamefont {Thomas}},
  \bibinfo {author} {\bibfnamefont {S.~K.}\ \bibnamefont {Kushwaha}}, \bibinfo
  {author} {\bibfnamefont {E.~D.}\ \bibnamefont {Bauer}}, \bibinfo {author}
  {\bibfnamefont {F.}~\bibnamefont {Ronning}}, \bibinfo {author} {\bibfnamefont
  {J.~D.}\ \bibnamefont {Thompson}}, \bibinfo {author} {\bibfnamefont
  {N.}~\bibnamefont {Harrison}},\ and\ \bibinfo {author} {\bibfnamefont
  {P.~F.~S.}\ \bibnamefont {Rosa}},\ }\bibfield  {title} {\bibinfo {title}
  {Ground state of {C}e$_{3}${B}i$_{4}${P}d$_{3}$ unraveled by hydrostatic
  pressure},\ }\href {https://doi.org/10.1103/PhysRevB.106.L161105} {\bibfield
  {journal} {\bibinfo  {journal} {Phys. Rev. B}\ }\textbf {\bibinfo {volume}
  {106}},\ \bibinfo {pages} {L161105} (\bibinfo {year} {2022})}\BibitemShut
  {NoStop}%
\bibitem [{\citenamefont {Mishra}\ \emph {et~al.}(2022)\citenamefont {Mishra},
  \citenamefont {Liu}, \citenamefont {Bauer}, \citenamefont {Ronning},\ and\
  \citenamefont {Thomas}}]{Mishra}%
  \BibitemOpen
  \bibfield  {author} {\bibinfo {author} {\bibfnamefont {S.}~\bibnamefont
  {Mishra}}, \bibinfo {author} {\bibfnamefont {Y.}~\bibnamefont {Liu}},
  \bibinfo {author} {\bibfnamefont {E.~D.}\ \bibnamefont {Bauer}}, \bibinfo
  {author} {\bibfnamefont {F.}~\bibnamefont {Ronning}},\ and\ \bibinfo {author}
  {\bibfnamefont {S.~M.}\ \bibnamefont {Thomas}},\ }\bibfield  {title}
  {\bibinfo {title} {Anisotropic magnetotransport properties of the
  heavy-fermion superconductor {C}e{R}h$_{2}${A}s$_{2}$},\ }\href
  {https://doi.org/10.1103/PhysRevB.106.L140502} {\bibfield  {journal}
  {\bibinfo  {journal} {Phys. Rev. B}\ }\textbf {\bibinfo {volume} {106}},\
  \bibinfo {pages} {L140502} (\bibinfo {year} {2022})}\BibitemShut {NoStop}%
\bibitem [{\citenamefont {Kadowaki}\ and\ \citenamefont
  {Woods}(1986)}]{Kadowaki}%
  \BibitemOpen
  \bibfield  {author} {\bibinfo {author} {\bibfnamefont {K.}~\bibnamefont
  {Kadowaki}}\ and\ \bibinfo {author} {\bibfnamefont {S.}~\bibnamefont
  {Woods}},\ }\bibfield  {title} {\bibinfo {title} {Universal relationship of
  the resistivity and specific heat in heavy-fermion compounds},\ }\href
  {https://doi.org/https://doi.org/10.1016/0038-1098(86)90785-4} {\bibfield
  {journal} {\bibinfo  {journal} {Solid State Communications}\ }\textbf
  {\bibinfo {volume} {58}},\ \bibinfo {pages} {507} (\bibinfo {year}
  {1986})}\BibitemShut {NoStop}%
\bibitem [{\citenamefont {Tsujii}\ \emph {et~al.}(2005)\citenamefont {Tsujii},
  \citenamefont {Kontani},\ and\ \citenamefont {Yoshimura}}]{Tsujii}%
  \BibitemOpen
  \bibfield  {author} {\bibinfo {author} {\bibfnamefont {N.}~\bibnamefont
  {Tsujii}}, \bibinfo {author} {\bibfnamefont {H.}~\bibnamefont {Kontani}},\
  and\ \bibinfo {author} {\bibfnamefont {K.}~\bibnamefont {Yoshimura}},\
  }\bibfield  {title} {\bibinfo {title} {Universality in heavy fermion systems
  with general degeneracy},\ }\href
  {https://doi.org/10.1103/PhysRevLett.94.057201} {\bibfield  {journal}
  {\bibinfo  {journal} {Phys. Rev. Lett.}\ }\textbf {\bibinfo {volume} {94}},\
  \bibinfo {pages} {057201} (\bibinfo {year} {2005})}\BibitemShut {NoStop}%
\bibitem [{\citenamefont {Jacko}\ \emph {et~al.}(2009)\citenamefont {Jacko},
  \citenamefont {Fjærestad},\ and\ \citenamefont {Powell}}]{Jacko}%
  \BibitemOpen
  \bibfield  {author} {\bibinfo {author} {\bibfnamefont {A.}~\bibnamefont
  {Jacko}}, \bibinfo {author} {\bibfnamefont {J.}~\bibnamefont {Fjærestad}},\
  and\ \bibinfo {author} {\bibfnamefont {B.}~\bibnamefont {Powell}},\
  }\bibfield  {title} {\bibinfo {title} {A unified explanation of the
  kadowaki–woods ratio in strongly correlated metals},\ }\href
  {https://doi.org/10.1038/nphys1249} {\bibfield  {journal} {\bibinfo
  {journal} {Nature Phys}\ }\textbf {\bibinfo {volume} {5}},\ \bibinfo {pages}
  {422–425} (\bibinfo {year} {2009})}\BibitemShut {NoStop}%
\end{thebibliography}%
\end{document}